\documentclass[reprint, aps, pra, letterpaper, superscriptaddress]{revtex4-2}%

\usepackage[T1]{fontenc} 

\usepackage{amsmath,color}
\usepackage{amssymb}
\usepackage{amsfonts}
\usepackage{amsthm}
\usepackage{mathtools}

\usepackage[title]{appendix}

\usepackage{pdfpages}

\usepackage{algorithm}
\usepackage{algpseudocode}
\usepackage{dsfont}

\usepackage{bbold}
\usepackage{chemformula}
\usepackage{siunitx}

\DeclarePairedDelimiterX\braket[2]{\langle}{\rangle}{#1 \delimsize\vert #2}
\DeclarePairedDelimiterX\braket3[3]{\langle}{\rangle}{#1 \delimsize\vert #2 \delimsize\vert #3}

\usepackage{accents}

\newcommand{\vE}{\mathbf{E}}

\newcommand{\ve}{\mathbf{e}}

\newcommand{\he}{\mathbf{e}}

\newcommand{\red}[1]{{\color{black} #1}}

\usepackage{etoolbox} 

\makeatletter
\patchcmd{\@outputpage@head}{\@ifx{\LS@rot\@undefined}{}{\LS@rot}}{}{}{}
\makeatother

\begin{document}

	\title{Energy-efficient pathway for selectively exciting solute molecules to high vibrational states via solvent vibration-polariton pumping}

	\author{Tao E. Li}%
	\email{taoli@sas.upenn.edu}
	
	\affiliation{Department of Chemistry, University of Pennsylvania, Philadelphia, Pennsylvania 19104, USA}
	\affiliation{Department of Chemistry, Yale University, New Haven, Connecticut, 06520, USA}

	\author{Abraham Nitzan} 
	\email{anitzan@sas.upenn.edu}
	\affiliation{Department of Chemistry, University of Pennsylvania, Philadelphia, Pennsylvania 19104, USA}
	\affiliation{School of Chemistry, Tel Aviv University, Tel Aviv 69978, Israel}

	\author{Joseph E. Subotnik}
	\email{subotnik@sas.upenn.edu}
	\affiliation{Department of Chemistry, University of Pennsylvania, Philadelphia, Pennsylvania 19104, USA}

	\begin{abstract}
	
	Selectively exciting target molecules to high vibrational states is inefficient in the liquid phase, which restricts the use of IR pumping to catalyze ground-state chemical reactions. Here, we demonstrate that this inefficiency can sometimes be solved by confining the liquid to an optical cavity under vibrational strong coupling conditions. For a liquid solution of \ch{$^{13}$CO2} solute in a \ch{$^{12}$CO2} solvent, cavity molecular dynamics simulations show that exciting a polariton (hybrid light-matter state) of the solvent with an intense laser pulse, under suitable resonant conditions, may lead to a very strong (> 3 quanta) and ultrafast (< 1 ps) excitation of the solute, even though the solvent ends up being barely excited. By contrast, outside a cavity the same input pulse fluence can  excite the solute by only half a vibrational quantum and the selectivity of excitation is low.  Our finding is robust under different cavity volumes, which may lead to observable cavity enhancement on IR photochemical reactions in Fabry--P\'erot cavities.
	\end{abstract}
	
	\maketitle
	
	\section*{Introduction}\label{sec:intro}
	
 	Controlling (electronic) ground-state chemical reaction rates with an external infrared (IR) laser  is a long-standing goal of IR photochemistry. Although it is possible to
 	trigger chemical reactions by exciting the reactants to high vibrational excited states  via  ladder climbing \cite{Maas1998}, such IR-controlled chemical reactions are rarely efficient in the liquid phase because of competing relaxation processes that
 	transfer vibrational energy to other degrees of freedom, either internal modes of the molecule or external modes  of surrounding molecules (leading to a build up of  heat in the environment) \cite{Owrutsky1994}.  
 	Up to now, there have been only a few observations of selective IR photochemistry (i.e. reactions where an IR pump promotes selective reactions beyond a simple heating effect) \cite{Delor2014Science,Stensitzki2018,Heyne2019}.  
 	That being said,  if there were a general mechanism for targeting some molecules (e.g., reactants or solute molecules)  and achieving highly vibrationally excited states on a timescale shorter than the lifetime of vibrational energy relaxation, one could imagine using IR light to help catalyze reactions of interest.
 	
 	During the past decade, a novel means to control molecular properties has emerged through the formation of  strong light-matter coupling between a vacuum electromagnetic field and a molecular (electronic or vibrational) transition of a large ensemble of molecules \cite{Herrera2019,Xiang2021JCP,Garcia-Vidal2021,Li2022Review}. Under this collective strong light-matter coupling regime, the formed hybrid light-matter states, known as molecular polaritons, can suppress ultraviolet-visible (UV-vis) photochemical reactions  \cite{Hutchison2012},  alter ground-state reaction rates \cite{Thomas2016,Thomas2019_science,Imperatore2021},  promote electronic conductivity \cite{Orgiu2015}, and facilitate energy transfer between different molecular species  \cite{Coles2014,Zhong2017,Saez-Blazquez2018,Du2018,Georgiou2021, Xiang2020Science}, etc. For example, under vibrational strong coupling (VSC) \cite{Shalabney2015,Long2015} between an optical cavity mode and an equal liquid mixture of \ch{W($^{12}$CO)6} and \ch{W($^{13}$CO)6} , Xiang \textit{et al} \cite{Xiang2020Science} have experimentally demonstrated that, by pumping the upper vibrational polariton (UP) with an external IR laser,  intermolecular vibrational energy transfer   between the two molecular species can be accelerated to a lifetime of tens of ps due to the intrinsically delocalized nature of polaritons.  
 	
 	Apart from the acceleration of  energy transfer due to polariton delocalization, the other intriguing consequence of VSC  is the enhancement of molecular nonlinear absorption \cite{Xiang2019State,Xiang2021JCP,Li2020Nonlinear,Ribeiro2020}. Due to the anharmonicity of molecular vibrations, it is possible to tune a cavity in such a way that  twice the lower polariton (LP) energy roughly matches the molecular $0\rightarrow 2$ vibrational transition; in such a case, pumping the LP can facilitate molecular nonlinear absorption  and promote the creation of highly vibrationally excited molecules \cite{Xiang2021JCP,Li2020Nonlinear}.

	Here, we will numerically demonstrate that by combining the above two unique features of VSC --- polariton delocalization and molecular anharmonicity --- it is possible to selectively excite a set of molecules, i.e., a small concentration of solute molecules,  to highly vibrationally excited states and leave the vast majority of solvent molecules barely excited via strongly exciting the solvent LP. We emphasize that this finding is of crucial importance, as it lays the foundations for using IR polaritonic chemistry to promote chemical reaction rates. Moreover, because the solvent LP is essentially composed of only solvent modes plus the cavity mode (with a negligible contribution from solute molecules),  the observation of energy accumulation in the solute molecules must be regarded as  vibrational energy transfer from solvent to solute, which is entropically unfavorable outside the cavity. The difference between our finding and conventional multiphoton molecular nonlinear absorption is also discussed in \red{Supplementary Figures 2, 3, and 8.}

	\begin{figure}
		\centering
		\includegraphics[width=1.0\linewidth]{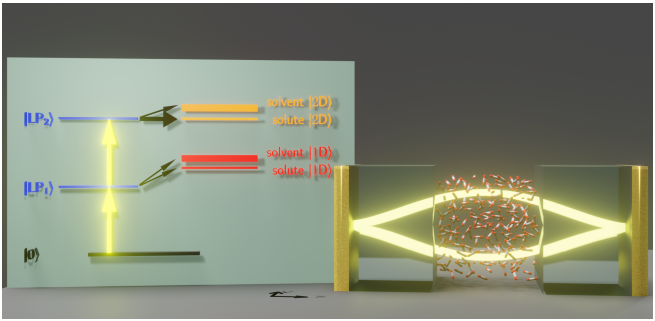}
		\caption{Sketch of the cavity setup for CavMD simulations. $N_{\text{sub}}$ molecules are confined in a cavity formed by a pair of parallel metallic  mirrors. These molecules are coupled to a single cavity mode (with two polarization directions) and are simulated in a periodic simulation cell. Our classical CavMD simulations show that, after a strong excitation of the solvent lower polariton (LP) with an IR laser pulse, the input energy can be selectively transferred to the highly excited dark states of the solute molecules, leaving the vast majority of the solvent molecules barely excited. The left cartoon demonstrates the quantum analog of our classical finding; see \red{Supplementary Note 8} for details.}
		\label{fig:toc}
	\end{figure}

	While many theoretical approaches \cite{Campos-Gonzalez-Angulo2020,LiHuo2020,Li2020Water,Sidler2021,Flick2017,Haugland2020,Triana2020Shape, Luk2017} have been developed  recently in order to describe molecular polaritons, our numerical approach is classical cavity molecular dynamics (CavMD) simulation \cite{Li2020Water,Li2020Nonlinear,Li2021Collective}, a new and promising tool for propagating  coupled photon-nuclear dynamics in the condensed phase, though with the limitations of using a classical force field to describe realistic molecules; see Fig. \ref{fig:toc} for the simulation setup and Methods (and also \red{Supplementary Methods}) for more details.

	\section*{Results}

		\begin{figure*}[!]
		\centering
		\includegraphics[width=1.0\linewidth]{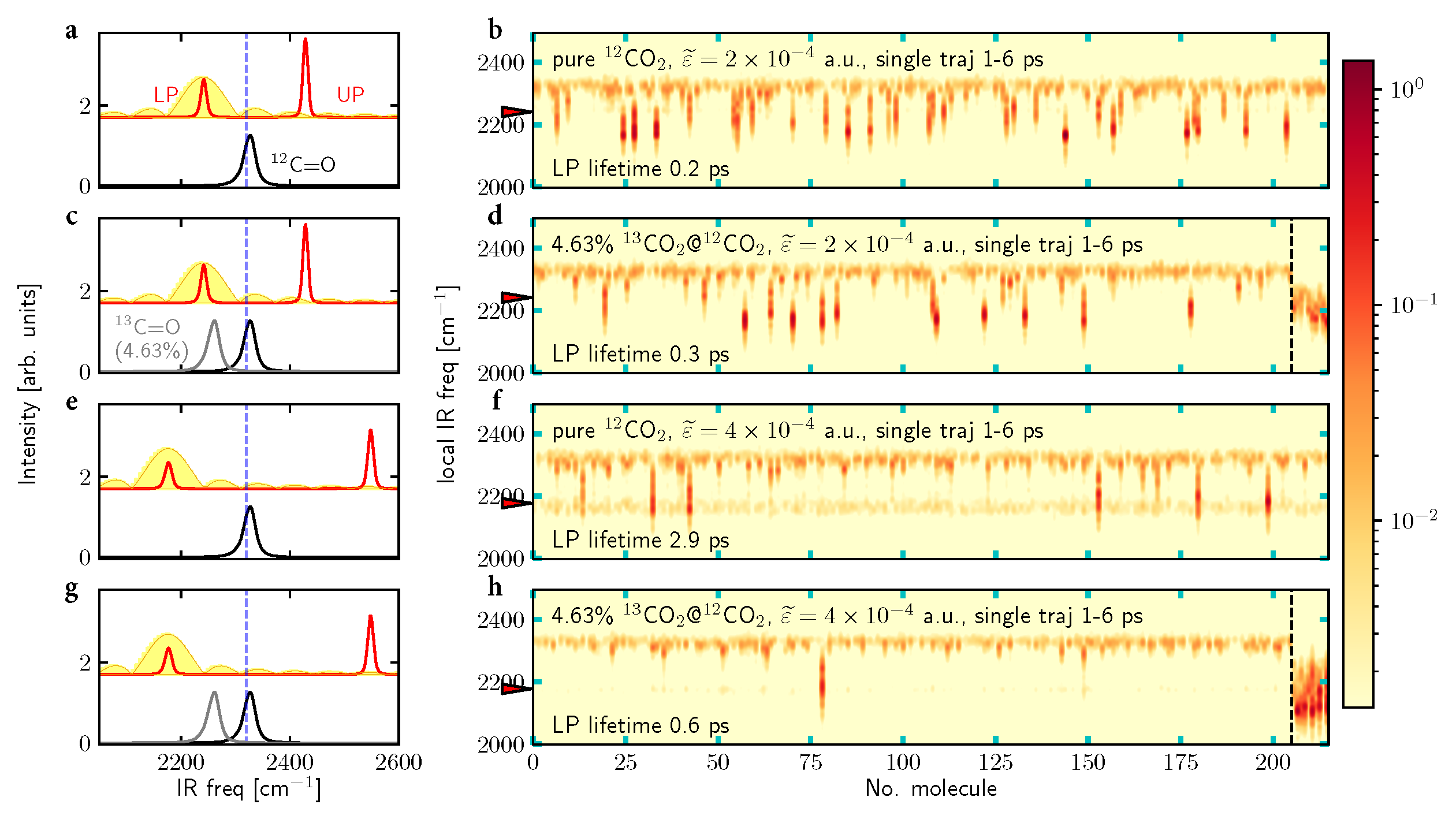}
		\caption{Rabi splitting and selective polariton energy transfer. (a) Equilibrium IR spectrum for pure liquid \ch{$^{12}$CO2} outside a cavity (black line) or inside  a cavity with \red{coupling} $\widetilde{\varepsilon}=2\times 10^{-4}$ a.u. (red line); the vertical dashed  blue line denotes the cavity mode at 2320 cm$^{-1}$. (b) The corresponding nonequilibrium local IR spectrum for each molecule after a resonant excitation of the LP at 2241 cm$^{-1}$ with a strong external pulse. The form of the external pulse takes $E_0\cos(\omega t + \phi)\he_x$ ($0.1 < t < 0.6$ ps) with fluence 632 mJ/cm$^{2}$; the corresponding pulse spectrum is also plotted as the yellow curve on the left panel. The nonequilibrium spectrum is calculated by evaluating the dipole auto-correlation function of each \ch{CO2} molecule over time interval 1-6 ps. The red arrow marks the frequency of the LP and also the external pulse, and the LP lifetime (0.2 ps, calculated from photonic energy dynamics \cite{Li2020Nonlinear}) is labeled at the bottom. (c) Equilibrium IR spectrum (red line) for 4.63\% \ch{$^{13}$CO2} in a \ch{$^{12}$CO2}  solution when $\widetilde{\varepsilon}=2\times 10^{-4}$ a.u. The gray line denotes the  \ch{$^{13}$CO2} IR spectrum outside the cavity; again, the black line denotes the pure  \ch{$^{12}$CO2} IR spectrum outside the cavity. (d) The corresponding nonequilibrium local IR spectrum for each molecule plotted in a similar manner as Fig. b. The responses of \ch{$^{12}$CO2} (left) and \ch{$^{13}$CO2} (right) are separated by the vertical dashed line. (e-h) The same plots as in  Figs. a-d  but with a larger effective light-matter coupling ($\widetilde{\varepsilon}=4\times 10^{-4}$ a.u.). Note that with an appropriate LP frequency, the LP  is more likely to transfer energy to the \ch{$^{13}$CO2} solute molecules than to the \ch{$^{12}$CO2} solvent. See \red{Supplementary Methods} for  simulation details, the definition of the \ch{CO2} force field, and the definition of the spectra signals.}
		\label{fig:excitation_statistics_solvent}
	\end{figure*}

	\subsection*{Polaritonic energy transfer in pure liquid}
	
	Before presenting the main finding of this manuscript --- selective energy transfer to solute molecules --- we investigate how exciting a polariton leads to a  nonuniform energy redistribution in a pure liquid \ch{$^{12}$CO2} system (with $N_{\text{sub}} = 216$ molecules explicitly propagated in a simulation cell) at room temperature (300 K). In Fig. \ref{fig:excitation_statistics_solvent}a, we  plot the equilibrium IR spectrum  outside a cavity (black line) or inside a cavity (red line), where the IR spectrum is calculated by evaluating the  auto-correlation function of the total dipole moment for the molecular subsystem \cite{Gaigeot2003,Li2020Nonlinear}. Inside the cavity, the \ch{C=O} asymmetric stretch mode (peaked at $\omega_0 = 2327$ cm$^{-1}$) is resonantly coupled to a cavity mode (at  $\omega_c = 2320$ cm$^{-1}$ corresponding to the vertical blue line) with an effective coupling strength $\widetilde{\varepsilon} = 2\times 10^{-4}$ a.u. In this collective VSC regime, a pair of polaritons [LP ($\omega_{\text{LP}} = 2241$ cm$^{-1}$) and UP ($\omega_{\text{UP}} = 2428$ cm$^{-1}$)] forms and these two peaks are separated by a Rabi splitting of 187 cm$^{-1}$. 
	
	We now excite the LP with a strong pulse of the form $\vE(t) = E_0\cos(\omega t) \ve_x$ ($0.1 < t < 0.6$ ps) with fluence 632 mJ/cm$^2$ ($E_0$ = $6\times 10^{-3}$ a.u.); see the yellow curve in Fig. \ref{fig:excitation_statistics_solvent}a for the corresponding pulse spectrum. This fluence will be used below and throughout the present manuscript, and the pulse is assumed to interact only with the molecular subsystem. Note that, as shown in \red{Supplementary Figure 6}, for the same amount of pumped energy, using a Gaussian pulse (instead of a square pulse) does not meaningfully alter the reported results; similarly, \red{as shown in Supplementary Figure 5}, if we assume that the external pulse activates the cavity mode directly (rather than the molecular subsystem), the reported results are effectively unchanged.
	
	In Fig. \ref{fig:excitation_statistics_solvent}b, after the LP pumping, the resulting energy distribution \red{among molecules}  is characterized by calculating the response of every \ch{$^{12}$CO2} molecule in the simulation cell after the LP excitation. The individual molecular response is calculated by evaluating the auto-correlation function of each single-molecule dipole moment (which we call the "local" IR spectrum) \cite{Li2020Nonlinear} during a time window $1 < t < 6$ ps (immediately after the LP excitation). Note that the response measured is different from  the usual IR spectrum --- which  is calculated by evaluating the auto-correlation function of the total dipole moment of a molecular system (as in Fig. \ref{fig:excitation_statistics_solvent}a) and reflects the dynamics of the molecular bright mode; the local IR spectrum measured here reflects the dynamics of individual molecules and can be expressed as a linear combination of molecular bright and dark modes. Because the total number of molecules explicitly accounted for in the simulation is large (with $N_{\text{sub}} = 216$), the local IR spectrum  is dominated by the density of dark modes.
	
	In  Fig. \ref{fig:excitation_statistics_solvent}b, every column of pixels represents the local IR spectrum  of one molecule (in total  $N_{\text{sub}} = 216$ molecules) during 1-6 ps and the colorbar (from light yellow to deep red) denotes a logarithmic scaled spectroscopic intensity. Clearly,  after the LP excitation (where the  LP lifetime fitted  from photonic energy dynamics \cite{Li2020Nonlinear} is 0.2 ps), the polaritonic energy is nonuniformly transferred to different \ch{$^{12}$CO2} molecules (which are composed mostly of vibrational dark modes). While most molecules are only weakly excited and have a peak near 2320 cm$^{-1}$ (the equilibrium IR peak; see also Fig. \ref{fig:excitation_statistics_solvent}a), a dozen of molecules are strongly excited and show an intense, red-shifted peak near 2200 cm$^{-1}$. Because we use an anharmonic force field \cite{Li2020Nonlinear} to simulate carbon dioxide molecules, the presence of intense, red-shifted peaks implies that the LP energy is mostly transferred to a small fraction of \ch{$^{12}$CO2} molecules. 

	The nonuniform polaritonic energy redistribution in Fig. \ref{fig:excitation_statistics_solvent}b stems from polariton-enhanced molecular nonlinear absorption, a mechanism which has been shown  experimentally (via two-dimensional IR spectroscopy) \cite{Xiang2019State}, analytically \cite{Ribeiro2020}, and numerically (via CavMD) \cite{Li2020Nonlinear}. Quantum-mechanically speaking, when twice the LP energy roughly matches the $0\rightarrow 2$ vibrational transition of molecules, the LP can serve as a "virtual state" to enhance molecular nonlinear absorption of light and directly create  highly vibrational excited molecules within a sub-ps timescale. In the present case (see Fig. \ref{fig:excitation_statistics_solvent}a), the LP frequency sits between the fundamental \ch{$^{12}$C=O} asymmetric stretch ($\omega_0 = 2327$ cm$^{-1}$) and the $1\rightarrow 2$ vibrational transition (which is roughly 2200 cm$^{-1}$). Hence, polariton-enhanced molecular nonlinear absorption can occur once the LP is strongly excited. Moreover, since molecules have different orientations and different local electrostatic environments (which leads to different instantaneous vibrational frequencies), only a small subset of molecules can interact strongly with the LP through the nonlinear channel (which requires an exact frequency match), thus leading to a vast difference in energy redistribution among molecules. In the end, some solvent molecules become highly excited and others do not. This quantum-mechanical interpretation of polariton-enhanced molecular nonlinear absorption is easy to understand. A classical analog can be pictured as well. Once the LP is excited and starts to dephase, due to the inhomogeneous local molecular environment, some molecules receive more polaritonic energy than the others. Due to molecular anharmonicity, these high-energy molecules oscillate with red-shifted, closer-to-LP frequencies, leading to an even stronger interaction between these high-energy molecules and the LP. Such "self-catalyzing" can, under strong LP excitation, eventually lead to some molecules being strongly and nonlinearly excited. While keeping in mind that our calculations are in fact classical, below we will use the quantum description because we feel the latter is more standard in the literature.

	\subsection*{Polaritonic energy transfer to solute molecules}
	
	Let us now show that the mechanism above can be utilized to selectively transfer the energy from an IR laser pulse to a few target molecules in a liquid system. For the sake of simplicity, let us choose these target molecules to be a few \ch{$^{13}$CO2} molecules and investigate how to achieve selective energy transfer to these \ch{$^{13}$CO2} solute molecules (which are dissolved in a liquid \ch{$^{12}$CO2} solution).
	
	For a solution of 4.63 \% (10/216) \ch{$^{13}$CO2} molecules dissolved in  liquid \ch{$^{12}$CO2} (in total $N_{\text{sub}} = 216$ molecules are included in the simulation cell), the red line in Fig. \ref{fig:excitation_statistics_solvent}c  plots the equilibrium IR spectrum  inside the same cavity as in Figs. \ref{fig:excitation_statistics_solvent}a,b (with $\omega_c = 2320$ cm$^{-1}$ and $\widetilde{\varepsilon} = 2\times 10^{-4}$ a.u.). Because the concentration of the \ch{$^{13}$CO2} solute  is small, inside the cavity, the positions of the polaritons are largely unchanged compared with the case of pure liquid \ch{$^{12}$CO2} (in  Fig. \ref{fig:excitation_statistics_solvent}a). Note that the equilibrium IR spectrum for pure liquid \ch{$^{13}$CO2} system outside a cavity is also plotted as the gray line in Fig. \ref{fig:excitation_statistics_solvent}c, where the \ch{$^{13}$C=O} asymmetric stretch peaks at 2262 cm$^{-1}$.
	
	For this liquid mixture, after a strong excitation of the LP (again with a pulse fluence of 632 mJ/cm$^2$), Fig. \ref{fig:excitation_statistics_solvent}d plots the transient local IR spectra for every \ch{$^{12}$CO2} and  \ch{$^{13}$CO2} molecule. Here, the  \ch{$^{12}$CO2} are plotted on the left and the \ch{$^{13}$CO2} molecules are plotted on the right hand side of the vertical dashed line. Fig. \ref{fig:excitation_statistics_solvent}d shows that there is now a competition between exciting \ch{$^{12}$CO2} and \ch{$^{13}$CO2} molecules, as molecules of both isotopes exhibit intense, red-shifted peaks.

\begin{figure*}
		\centering
		\includegraphics[width=1.0\linewidth]{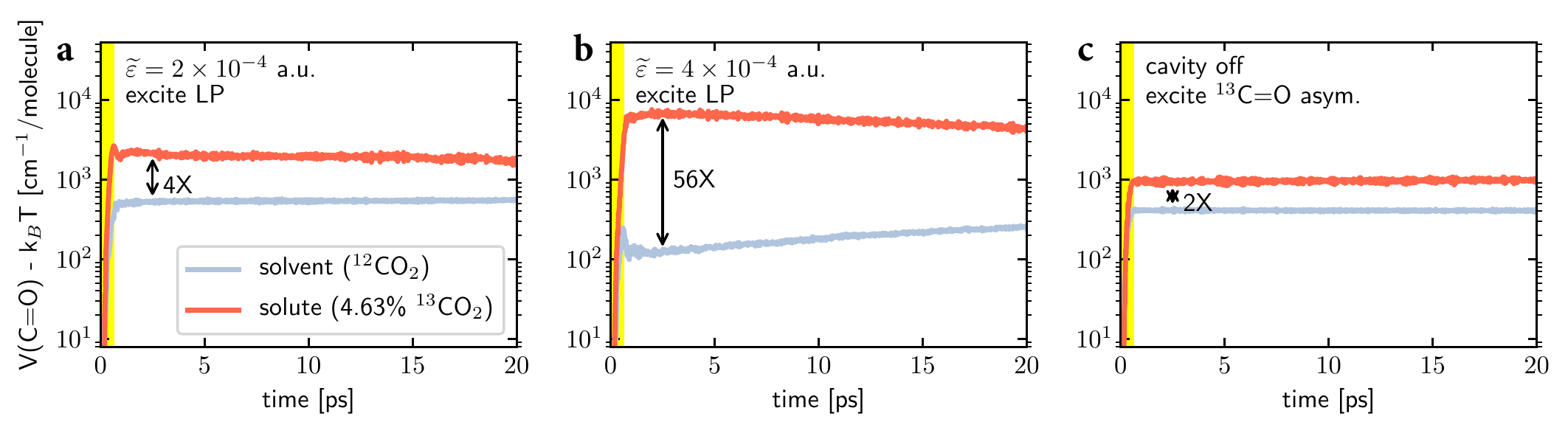}
		\caption{Time-resolved dynamics for the average \ch{C=O} bond potential energy per molecule. Three conditions are compared: (a) $\widetilde{\varepsilon} = 2\times 10^{-4}$ a.u. after exciting the LP at 2241 cm$^{-1}$ (the same as Fig. \ref{fig:excitation_statistics_solvent}d); (b) $\widetilde{\varepsilon} = 4\times 10^{-4}$ a.u. after exciting the LP  at 2177 cm$^{-1}$ (the same as Fig. \ref{fig:excitation_statistics_solvent}h); and (c) outside the cavity after  exciting the \ch{$^{13}$C=O} asymmetric stretch  at 2262 cm$^{-1}$. In  each subplot, the \ch{$^{12}$CO2} solvent is labeled with gray line and the 4.63\% \ch{$^{13}$CO2} solute is labeled with red line. The  $y$-axis  is plotted in a logarithmic scale, and the yellow region denotes the time window ($0.1 < t < 0.6$ ps) during which the external pulse is applied. Note that although the nonequilibrium  local IR spectrum in Fig. \ref{fig:excitation_statistics_solvent} was calculated using a single trajectory,  40 trajectories are averaged here to obtain an ensemble-averaged result; see \red{Supplementary Methods for simulation details.} }
		\label{fig:VCO_param}
	\end{figure*}

	This conclusion is illustrated quantitatively in Fig. \ref{fig:VCO_param}a, where we plot the dynamics of the average \ch{C=O} bond potential energy  (minus the thermal energy $k_BT$) per \ch{$^{12}$CO2} (gray line) or \ch{$^{13}$CO2} (red line) molecule after the strong LP excitation.  Due to the LP excitation [where the time window of the pulse ($0.1 < t < 0.6$ ps) is represented with a yellow region], the vibrational energy  of the \ch{$^{13}$CO2} solute molecules can reach roughly four times that of the \ch{$^{12}$CO2}  solvent molecules and each solute molecule can absorb roughly a vibrational quantum ($\sim 2\times 10^3$ cm$^{-1}$) of the input energy.  Note that Fig. \ref{fig:VCO_param}a is drawn with a logarithmic scale; see the black arrow which denotes the energy difference. 
	By contrast, outside the cavity, when the \ch{$^{13}$C=O} asymmetric stretch of the liquid mixture is resonantly excited by the same pulse (Fig. \ref{fig:VCO_param}c),   the \ch{$^{13}$CO2} vibrational energy reaches roughly twice the energy of the \ch{$^{12}$CO2} solvent molecules,  and the energy absorption of each solute molecule is roughly half a vibrational quantum ($\sim 10^3$ cm$^{-1}$). Our finding indicates that the cavity environment can meaningfully improve the selectivity and also the absolute value of solute excitation by a factor of two (though perhaps not a huge effect).

	\subsection*{Improving the selectivity of energy transfer  to solute molecules by tuning the Rabi splitting}
	
	In order to improve the selectivity of energy transfer from the LP, we need to both (i) suppress the energy transfer to the \ch{$^{12}$CO2}  solvent molecules and also (ii) enhance the energy transfer to the \ch{$^{13}$CO2}  solute molecules. As mentioned above,  the nonuniform polaritonic energy transfer stems from polariton-enhanced molecular nonlinear absorption, a mechanism which requires twice the LP frequency to roughly match the $0\rightarrow 2$ vibrational transition. In other words, by changing the LP frequency relative to the vibration of one molecular species, we can enhance or suppress  polaritonic energy transfer to that molecular species according to a frequency match or mismatch.

	Let us first focus on the side of the \ch{$^{12}$CO2}  molecules and investigate how to suppress the energy transfer to the \ch{$^{12}$CO2} molecules with an increased Rabi splitting.
	For a pure liquid \ch{$^{12}$CO2} system, when  the effective light-matter coupling strength is increased from $\widetilde{\varepsilon} = 2\times 10^{-4}$ a.u. to $\widetilde{\varepsilon} = 4\times 10^{-4}$ a.u., the red line in Fig. \ref{fig:excitation_statistics_solvent}e  plots the equilibrium IR spectrum inside the  cavity with $\omega_c = 2320$ cm$^{-1}$. For the larger light-matter coupling ($\widetilde{\varepsilon} = 4\times 10^{-4}$ a.u.), the Rabi splitting increases by a factor of two and the LP frequency red-shifts from 2241 cm $^{-1}$ (in Figs. \ref{fig:excitation_statistics_solvent}a-d) to 2177 cm$^{-1}$. Now, as shown in Fig. \ref{fig:excitation_statistics_solvent}f, after a resonant excitation of the LP with a strong laser pulse,  polaritonic energy transfer to  \ch{$^{12}$CO2} molecules is largely suppressed compared with Fig. \ref{fig:excitation_statistics_solvent}b. The suppression of polaritonic energy transfer is consistent with a longer LP lifetime (2.9 ps) relative to that in Fig. \ref{fig:excitation_statistics_solvent}b (0.2 ps); this longer LP lifetime can be observed in Fig. \ref{fig:excitation_statistics_solvent}f by visualizing the near uniform intensity among all molecules at the LP frequency (positioned by the red arrow)  over the time scale 1-6 ps, which corresponds to a long lived  molecular bright state.

	Second, consider the case where the liquid mixture of 4.63 \% \ch{$^{13}$CO2} in a \ch{$^{12}$CO2} solution experiences a larger light-matter coupling (with $\widetilde{\varepsilon} = 4\times 10^{-4}$ a.u.). Because the concentration of the \ch{$^{13}$CO2} solute molecules is small, the equilibrium IR spectrum inside the cavity (Fig. \ref{fig:excitation_statistics_solvent}g) remains unchanged compared with the case of  pure liquid \ch{$^{12}$CO2} (Fig. \ref{fig:excitation_statistics_solvent}e). After resonantly exciting the LP with a strong laser pulse, as shown in Fig. \ref{fig:excitation_statistics_solvent}h, we find that the polaritonic energy  mostly transfers to the \ch{$^{13}$CO2}  molecules (on the right hand side of the vertical dashed white line). Quantitatively speaking, Fig. \ref{fig:VCO_param}b plots the corresponding dynamics of the average \ch{C=O} bond potential energy for the solute and solvent molecules. Here, within the time window of the laser pulse ($t < 1$ ps), the solute molecules can be excited to a state of roughly three vibrational quanta (recall that the frequency of the \ch{C=O} asymmetric stretch is roughly $\sim 2\times 10^3$ cm$^{-1}$), which is six times of the outside-cavity solute energy absorption (Fig. \ref{fig:VCO_param}c). As far as  selectivity is considered, within $t < 5$ ps, the solute molecules  absorb an energy 56 times larger than  the solvent molecules (see the black arrow), a difference which is one-order-of-magnitude larger than the case of outside the cavity (Fig. \ref{fig:VCO_param}c). In short, although we found above that,  when $\widetilde{\varepsilon} = 2\times 10^{-4}$ a.u., there is a competition between the polaritonic energy transfer to  \ch{$^{12}$CO2} and \ch{$^{13}$CO2} molecules, we now find that when  $\widetilde{\varepsilon} = 4\times 10^{-4}$ a.u., \ch{$^{13}$CO2} clearly wins.

	The high selectivity of this polaritonic energy transfer to the solute molecules can be explained as follows.  First, since the LP frequency is unchanged compared with the pure liquid \ch{$^{12}$CO2} system in Fig. \ref{fig:excitation_statistics_solvent}e, polariton-enhanced molecular nonlinear absorption  for the \ch{$^{12}$CO2} solvent molecules remains greatly suppressed, which is analogous to Fig. \ref{fig:excitation_statistics_solvent}f.  Second, for the larger Rabi splitting, the vibrational frequency difference between the LP and \ch{$^{13}$C=O} asymmetric stretch (see red and gray lines in Fig. \ref{fig:excitation_statistics_solvent}g) is suitable to allow for polariton-enhanced molecular nonlinear absorption of the solute molecules. This statement is consistent with the shortened LP lifetime in Fig. \ref{fig:excitation_statistics_solvent}h (0.6 ps) as compared with $\tau_{\text{LP}} = 2.9$ ps in Fig. \ref{fig:excitation_statistics_solvent}f.
	All together, these two facts lead to a high selectivity in  energy transfer. A summary of the quantum picture of our finding --- a selective polariton energy transfer to the second excited solute dark states when twice the solvent LP frequency roughly matches the solute $0\rightarrow 2$ transition --- is  plotted in cartoon form on the left hand side of  Fig. \ref{fig:toc}; see also \red{Supplementary Note 8} for details.

	\subsection*{Dependence on system size and solute concentration}

	\begin{figure}[h]
	\centering
	\includegraphics[width=1.0\linewidth]{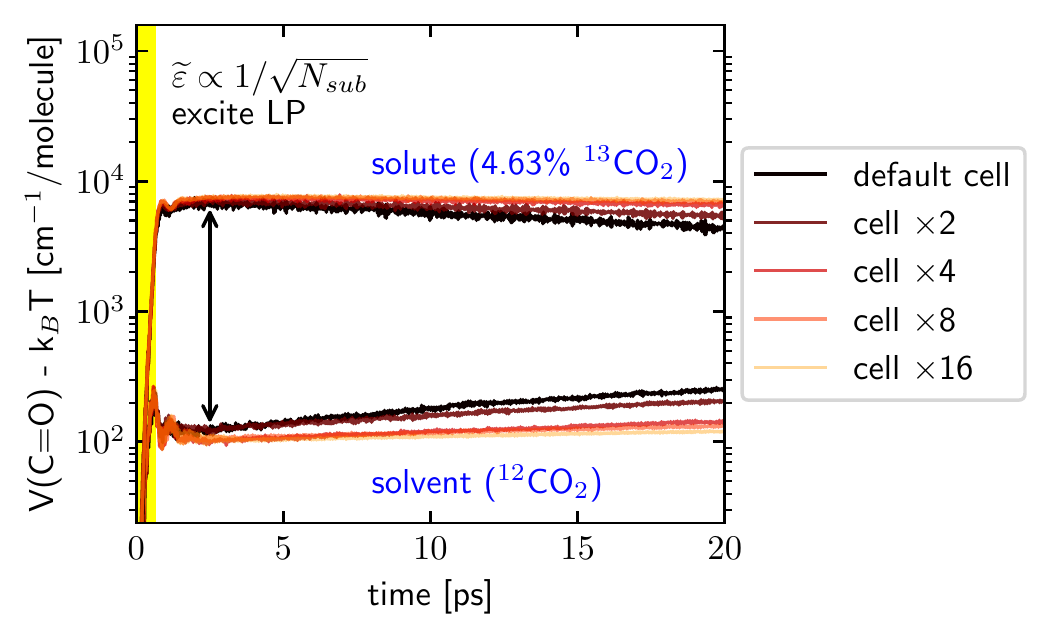}
	\caption{Effects of the molecular system size on the dynamics of \ch{C=O} bond potential energy \red{at constant solute concentration. The Rabi splitting and molecular number density have been fixed.} Black lines correspond to the data in Fig. \ref{fig:VCO_param}b ($N_{\text{sub}} = 216$), and the lines  from dark red to orange represent increasing molecular systems where $N_{\text{sub}}$ is increased by a factor of $N$ ($N=2,4,8,16$).  In all cases, we excite the LP and measure the amount of potential energy in the \ch{C=O} bonds. Note that the selective polaritonic energy transfer to the \ch{$^{13}$CO2} solute molecules is fairly robust against molecular system size, indicating that the current finding may hold for collective VSC with large cavity volumes but under the same Rabi splitting, i.e., smaller $\widetilde{\varepsilon}$; see text below \red{and also Supplementary Figure 1} for an explanation of the system size dependence.}
	\label{fig:VCO_pbc}
	\end{figure}

	One important issue we must address is the dependence of polaritonic energy transfer on molecular system size. This study is necessary because usual VSC setups in experiments (e.g., Fabry--P\'erot microcavities)  contain a macroscopic number of molecules, while we have only simulated a molecular system of $N_{\text{sub}} = 216$ coupled to  the cavity. Therefore,  we must check how our conclusions vis a vis VSC effects depend on molecular system size.
	
	Following the approach from Refs. \cite{Li2020Nonlinear}, in Fig. \ref{fig:VCO_param}b, we enlarge the molecular system by simultaneously increasing both the number of  \ch{$^{13}$CO2} and  \ch{$^{12}$CO2} molecules (maintaining a constant proportion) and we decrease the light-matter coupling strength ($\widetilde{\varepsilon}$); these two effects are chosen in a balanced manner ($\widetilde{\varepsilon} \propto 1/\sqrt{N_{\text{sub}}}$) so as to keep the molecular density and Rabi splitting the same as in Figs. \ref{fig:excitation_statistics_solvent}g,h. For systems with different sizes, after the same strong excitation of the LP, Fig. \ref{fig:VCO_pbc} plots the dynamics of the average \ch{C=O} bond potential energy for  the \ch{$^{13}$CO2} and  \ch{$^{12}$CO2} molecules. Here, the top and bottom lines denote the dynamics of \ch{$^{13}$CO2} and  \ch{$^{12}$CO2} molecules, respectively. Lines with different colors (from black to orange) denote  molecular systems of different sizes (from $N_{\text{sub}} = 216$ to $N_{\text{sub}} = 216\times 16$). Note that after the pulse, the \ch{$^{13}$CO2}  molecules are hot and start to cool down, while the  weakly excited \ch{$^{12}$CO2}  molecules start to heat up. Fig. \ref{fig:VCO_pbc} also clearly shows that, although the later-time vibrational relaxation and energy transfer dynamics ($t > 5$ ps)  can depend on the size of the molecular system \cite{Li2021Collective}, we do observe a consistent selective polaritonic energy transfer behavior at early times ($t < 5$ ps)  for all molecular system sizes. Therefore, we tentatively conclude that our numerical findings here should be observable in cavities with different volumes, including  Fabry--P\'erot microcavities (that are usually used to study collective VSC) and plasmonic cavities, an emerging platform for studying VSC \cite{Brawley2021,Yoo2021}.
	
	Note that the robustness of the early-time (< 1 ps) polaritonic decay dynamics against the system size can be rationalized by, e.g., Fermi's golden rule calculations. Although the interaction between the polariton and each dark mode scales with $1/N$ (where $N$ denotes the total molecular number) via intermolecular interactions, since there are $N-1$ dark modes which allow polaritonic energy transfer (or dephasing), when Fermi's golden rule is used to describe the polaritonic relaxation dynamics, the $1/N$ and $N-1$ factors cancel with each other, leading to consistent early-time dynamics during polaritonic relaxation versus the system size. 
	This argument explains why the early time dynamics (< 1 ps) in Fig. \ref{fig:VCO_pbc} do not depend sensitively on $N$.  That being said, for the later-time (> 5 ps) dynamics, which involve energy transfer and vibrational relaxation, the same argument shows that the presence of a cavity will play a role and lead to $N$-dependent dynamics. After all, consider the following energy transfer channel that we might expect to be important after we have excited a set of \ch{$^{13}$CO2} molecules:  \ch{$^{13}$CO2} $\rightarrow$ polaritons $\rightarrow$  \ch{$^{12}$CO2}.  Since there are only two (not $N$) polaritons, we do not expect this pathway to play an important role once the system size is very large, leading to slower vibrational relaxation and energy transfer dynamics in later times; see Ref. \cite{Li2021Collective} for a detailed study.

		\begin{figure}[h]
		\centering
		\includegraphics[width=1.0\linewidth]{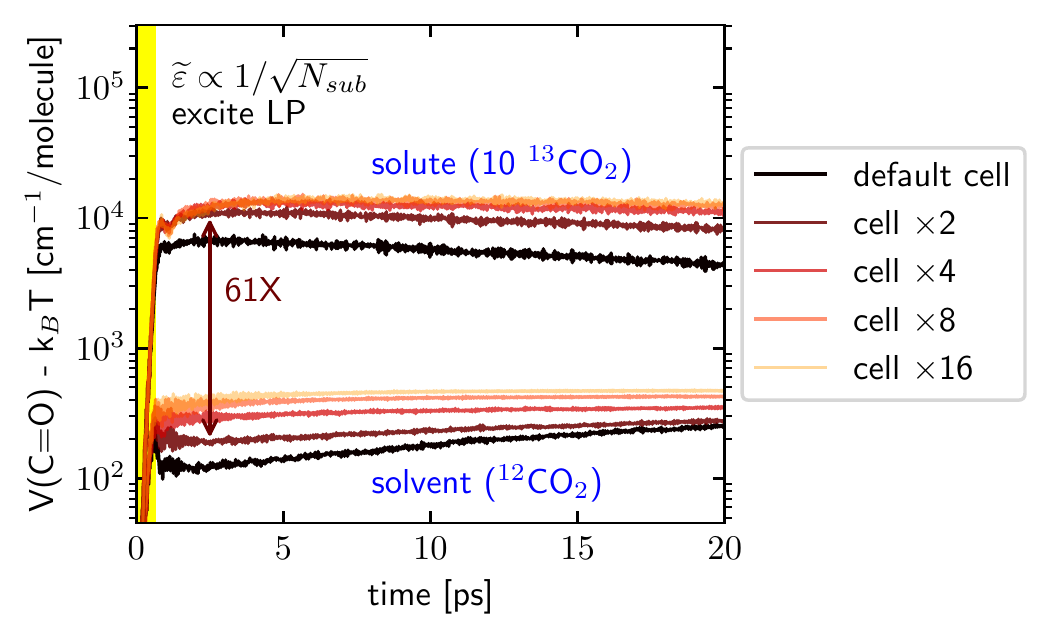}
		\caption{\red{Effects of the molecular system size on the dynamics of \ch{C=O} bond potential energy at constant solute molecular number. This plot is the same as Fig. \ref{fig:VCO_pbc} except for that the \ch{$^{13}$CO2} solute molecular number is fixed as 10 when the system size enlarges.} Note that, when  the system size is increased (or when the solute concentration is decreased), both the solvent and solute molecules are more heavily pumped under the same IR excitation of the LP. See text for details of this counter-intuitive feature.}
		\label{fig:VCO_pbc2}
	\end{figure}
	
	Finally, in Fig.  \ref{fig:VCO_pbc2}, we plot the dynamics of averaged \ch{C=O} bond potential energy per \ch{$^{13}$CO2} solute or \ch{$^{12}$CO2} solvent molecule as a function of time for different system sizes --- but now with a fixed number (10) of solute molecules. In other words, we increase only the number of \ch{$^{12}$CO2} solvent molecules (so that the solute concentration is decreased). For all curves, as above, the total Rabi splitting is kept constant by decreasing the effective coupling strength $\widetilde{\varepsilon}$, which corresponds physically to placing the system in a larger volume cavity.
	
	Perhaps counter-intuitively, Fig. \ref{fig:VCO_pbc2} shows that  the \ch{C=O} bond energy of both the solvent and solute molecules  increases as the number of solvent molecules increases. These findings can be explained as follows.  On the one hand, as for the solute molecules, when the number of solvent molecules increases, the total transition dipole of the solvent bright mode also increases, such that the total amount of energy absorbed through the polariton grows when the same incident pulse is applied --- and each solute molecule will recover more energy from the polariton through the nonlinear channel (since the energy absorbed by the polariton divided by the solute number grows). On the other hand, when the concentration of the solute decreases (under larger system sizes),  the pumped solvent polariton cannot efficiently transfer energy to its favorite acceptor (the solute), such that the polariton transfers more energy to the solvent molecules.  Finally, with regards to selectivity (i.e., energy of solute divided by energy of solvent), these two effects clearly work against each other. Empirically, we find that the selectivity reaches a maximum value of 61 times (see the vertical brown arrow) and the solute excitation can exceed $10^4$ cm$^{-1}$ (4$\sim$ 5 quanta) when we double the system size (i.e. we add slightly more than two times the number of solvent molecules, so that the solute concentration becomes  2.32\%, the brown lines). Very interestingly, in \red{Supplementary Figure 7} we  show that the selectivity also  goes through a maximum value as a function of the pulse fluence, confirming our assertion that energy transfer to the \ch{$^{13}$CO2}  species is dominated by a nonlinear transition: at low intensities this mechanism is inefficient, while at high intensities  the \ch{$^{13}$CO2}  species energy absorption becomes saturated.
	
	The findings above cannot be achieved outside a cavity. For example, on the one hand, when the solvent bright mode is pumped outside a cavity, since the excited bright mode dephases on a much faster timescale than the polaritons \cite{Grafton2020}, and because there is no resonance conditions between the solvent bright mode and the solute, energy transfer from the solvent to solute is not efficient and occurs on a much longer timescale than $1$ ps. On the other hand, when the solute vibration is directly excited outside the cavity (see Fig. \ref{fig:VCO_param}c), the energy absorbed by the system is much smaller than the inside-cavity case (where energy absorption is dominated by the solvent polariton).

	\section*{Discussion}

	To summarize, we have numerically demonstrated that energy from a strong IR laser pulse can be  selectively transferred to solute molecules by pumping the LP of the solvent molecules. This selective energy transfer behavior  requires  the LP to have an appropriate frequency to selectively enhance molecular nonlinear absorption for the solute (but not solvent) molecules and can be optimized by changing the solute concentration. Consequently, the transient vibrational energy difference between the solute and the solvent molecules can exceed the free-space case by more than one order of magnitude.  	Since our CavMD simulations have shown some key agreements with VSC experiments (including the signature of polariton-enhanced molecular nonlinear absorption), and because the current finding of selective energy transfer to the solute molecules is robust against the cavity volume,  we believe that this finding is also amenable to experimental verification.
	For example, although the molecule-resolved nonequilibrium local IR spectrum (see Fig. \ref{fig:excitation_statistics_solvent}) cannot be directly observed experimentally, for Fabry--P\'erot microcavities, pump-probe or 2D-IR spectroscopy should be able to verify the main predictions of this paper. For example, once a small fraction of solute molecules is added, the observation of a strong reduction of the solvent LP lifetime when twice the solvent LP frequency roughly matches the solute $0\rightarrow 2$ transition would provide one indirect piece of evidence supporting our mechanism. Moreover, for certain experiments, after pumping the solvent LP, the observation of  bond dissociation events would serve as a direct piece of evidence confirming that solute molecules have been highly excited.

	In principle, our finding might be one route to IR laser-controlled chemistry on an electronic ground-state surface, which remains difficult to achieve in the liquid phase. In order to selectively excite the (presumably chemically active) solute molecules in an efficient way, it would be best if the solvent molecules were to have a relatively simple vibrational structure so that the dephasing channel from the solvent polariton to the solvent dark modes can be largely forbidden. For the solute molecules, however, a complicated vibrational density of states might actually enhance the energy transfer from the solvent polariton. One must wonder if one potential application of the present manuscript would be isotope separation, where two similar molecules can be distinguished by a vibrational degree of freedom.  Finally, for the cases where the vibrational relaxation is so strong that the transient excitation of the solute molecules would not facilitate IR photochemistry, the current mechanism may also allow us to achieve a selective heating of a local geometry, e.g., by putting the solute molecules inside a protein, perhaps one can control a selective denaturation of proteins without heating up the whole liquid sample. Overall, there are clearly many future possibilities for using collective VSC to modify molecular properties.

	\section*{Methods}	
	
	The CavMD approach is formulated in the dipole gauge and includes the self-dipole term in the light-matter Hamiltonian, thus preserving gauge invariance \cite{Schafer2020,Taylor2020,Stokes2020}. The motivation for using classical CavMD for simulating nonequilibrium vibrational dynamics inside the cavity is twofold. First, for molecular vibrational relaxation and energy transfer dynamics in the condensed phase and outside the cavity, classical (nonequilibrium) molecular dynamics simulations have been shown to often agree well with experiments \cite{Heidelbach1998,Kabadi2004,Kandratsenka2009,Miller1978,Orel1980}; moreover,  cavity photons are harmonic oscillators, whose dynamics can usually be  captured well by classical dynamics \cite{Miller1978}. By combining the classical motion of cavity photons and molecules, CavMD is expected to capture the essential physics in many VSC experiments. Second,
	CavMD has already correctly captured some key features of both equilibrium and nonequilibrium VSC experiments \cite{Vergauwe2019,Xiang2018,Xiang2019State,Grafton2020}, including (i) the asymmetry of a Rabi splitting \cite{Li2020Water,Vergauwe2019},  (ii) polaritonic relaxation to vibrational dark modes on a timescale of ps or sub-ps \cite{Li2020Nonlinear,Xiang2018,Grafton2020},  (iii) polariton-enhanced molecular nonlinear absorption \cite{Li2020Nonlinear} --- a phenomenon whose signature is a delayed  population gain in the first excited state of vibrational dark modes after strongly exciting the LP \cite{Xiang2019State,Xiang2021JCP}, and (iv) dark-mode relaxation dynamics for systems with large cavity volumes \cite{Li2021Collective,Grafton2020}. Altogether, classical CavMD appears to be a suitable choice between accuracy and affordability for simulating nonequilibrium vibrational dynamics under VSC.
	
	The fundamental equations underlying CavMD and all simulation details for exciting a polariton in a liquid carbon dioxide system (including the force field) can be found in Refs. \cite{Li2020Water,Li2020Nonlinear}; see also \red{Supplementary Methods} for a comprehensive explanation. In short, as shown in Fig. \ref{fig:toc}, we imagine $N_{\text{sub}}$ molecules in a periodic simulation cell coupled to a single cavity mode (with two polarization directions). The effective coupling strength between each molecule and the cavity mode is denoted as $\widetilde{\varepsilon}$, which is defined as
	\begin{equation}
		\widetilde{\varepsilon} \equiv \sqrt{N_{\text{cell}}} \varepsilon .
	\end{equation}
	Here, $\varepsilon = \sqrt{m_c\omega_c^2/\Omega\epsilon_0}$ denotes the true light-matter coupling strength between each molecule and the cavity mode, where $\omega_c$, $\Omega$ and $\epsilon_0$ denote the cavity mode frequency, the effective cavity volume, and the vacuum permittivity, respectively, and $m_c = 1$ a.u. denotes the auxiliary mass for the cavity photons (see \red{Supplementary Methods} for details). $N_{\text{cell}}$ denotes the number of periodic simulation cells, the value of which can be determined by fitting the experimentally observed Rabi splitting. As discussed in \red{Supplementary Methods}, when  CavMD is used to simulate VSC in Fabry--P\'erot microcavities, we must always check that our results  on a small subsystem are not spurious.  We check for this possibility by running a protocol that enlarges the molecular system size (while fixing the Rabi splitting and molecular density) --- the results in the asymptotic limit of a very large $N_{\text{sub}}$ correspond to the true experimental result. See Fig. \ref{fig:VCO_pbc} for such a system size enlarging process.

	For the simulations reported here, by using an explicit force field defined in \red{Supplementary Methods}, we  include vibrational relaxation and dephasing for the molecular subsystem. We have assumed that the molecular force field remains the same inside versus outside the cavity, i.e., both intramolecular and intermolecular interactions are unchanged under VSC. For nonpolar molecules, although a recent first-principle calculation shows that intermolecular interactions can be modified by several cm$^{-1}$ when a few molecules are confined in a cavity \cite{Haugland2021}, we anticipate that, compared with thermal fluctuations at 300 K, such a small modification can be neglected when the molecular number is large. We disregard cavity loss in the manuscript, which should usually be reasonable for Fabry--P\'erot microcavities.  Note that the main finding of this manuscript --- selective polaritonic energy transfer to solute molecules --- occurs mostly within 1 ps after the laser pumping, while a typical cavity mode lifetime for a Fabry--P\'erot microcavity is $\sim 5$ ps \cite{Xiang2019State}.  In \red{Supplementary Note 2} we have verified that a cavity lifetime larger than 1 ps would not meaningfully alter the results corresponding to Fabry--P\'erot microcavities (i.e., when $N_{\text{sub}}$ is large). The code for reproducing this work, which is implemented based on the I-PI interference \cite{Kapil2019} and LAMMPS \cite{Plimpton1995}, is open-source available at Github \cite{TELi2020Github}.

	\section*{Data availability}
	All the data shown here, including the script files to generate the raw molecular dynamics trajectories  and the data for plotting, are available at \url{https://github.com/TaoELi/cavity-md-ipi} \cite{TELi2020Github}.
	
	\section*{Code availability}
	The code for reproducing this paper,  including a tutorial for installation and simulations, is available at \url{https://github.com/TaoELi/cavity-md-ipi} \cite{TELi2020Github}.


\begin{thebibliography}{52}%
		\makeatletter
		\providecommand \@ifxundefined [1]{%
			\@ifx{#1\undefined}
		}%
		\providecommand \@ifnum [1]{%
			\ifnum #1\expandafter \@firstoftwo
			\else \expandafter \@secondoftwo
			\fi
		}%
		\providecommand \@ifx [1]{%
			\ifx #1\expandafter \@firstoftwo
			\else \expandafter \@secondoftwo
			\fi
		}%
		\providecommand \natexlab [1]{#1}%
		\providecommand \enquote  [1]{``#1''}%
		\providecommand \bibnamefont  [1]{#1}%
		\providecommand \bibfnamefont [1]{#1}%
		\providecommand \citenamefont [1]{#1}%
		\providecommand \href@noop [0]{\@secondoftwo}%
		\providecommand \href [0]{\begingroup \@sanitize@url \@href}%
		\providecommand \@href[1]{\@@startlink{#1}\@@href}%
		\providecommand \@@href[1]{\endgroup#1\@@endlink}%
		\providecommand \@sanitize@url [0]{\catcode `\\12\catcode `\$12\catcode
			`\&12\catcode `\#12\catcode `\^12\catcode `\_12\catcode `\%12\relax}%
		\providecommand \@@startlink[1]{}%
		\providecommand \@@endlink[0]{}%
		\providecommand \url  [0]{\begingroup\@sanitize@url \@url }%
		\providecommand \@url [1]{\endgroup\@href {#1}{\urlprefix }}%
		\providecommand \urlprefix  [0]{URL }%
		\providecommand \Eprint [0]{\href }%
		\providecommand \doibase [0]{https://doi.org/}%
		\providecommand \selectlanguage [0]{\@gobble}%
		\providecommand \bibinfo  [0]{\@secondoftwo}%
		\providecommand \bibfield  [0]{\@secondoftwo}%
		\providecommand \translation [1]{[#1]}%
		\providecommand \BibitemOpen [0]{}%
		\providecommand \bibitemStop [0]{}%
		\providecommand \bibitemNoStop [0]{.\EOS\space}%
		\providecommand \EOS [0]{\spacefactor3000\relax}%
		\providecommand \BibitemShut  [1]{\csname bibitem#1\endcsname}%
		\let\auto@bib@innerbib\@empty
		\bibitem [{\citenamefont {Maas}\ \emph {et~al.}(1998)\citenamefont {Maas},
			\citenamefont {Duncan}, \citenamefont {Vrijen}, \citenamefont {van~der
				Zande},\ and\ \citenamefont {Noordam}}]{Maas1998}%
		\BibitemOpen
		\bibfield  {author} {\bibinfo {author} {\bibfnamefont {D.}~\bibnamefont
				{Maas}}, \bibinfo {author} {\bibfnamefont {D.}~\bibnamefont {Duncan}},
			\bibinfo {author} {\bibfnamefont {R.}~\bibnamefont {Vrijen}}, \bibinfo
			{author} {\bibfnamefont {W.}~\bibnamefont {van~der Zande}},\ and\ \bibinfo
			{author} {\bibfnamefont {L.}~\bibnamefont {Noordam}},\ }\bibfield  {title}
		{\bibinfo {title} {{Vibrational ladder climbing in NO by (sub)picosecond
					frequency-chirped infrared laser pulses}},\ }\href
		{https://doi.org/10.1016/S0009-2614(98)00531-4} {\bibfield  {journal}
			{\bibinfo  {journal} {Chem. Phys. Lett.}\ }\textbf {\bibinfo {volume}
				{290}},\ \bibinfo {pages} {75} (\bibinfo {year} {1998})}\BibitemShut
		{NoStop}%
		\bibitem [{\citenamefont {Owrutsky}\ \emph {et~al.}(1994)\citenamefont
			{Owrutsky}, \citenamefont {Raftery},\ and\ \citenamefont
			{Hochstrasser}}]{Owrutsky1994}%
		\BibitemOpen
		\bibfield  {author} {\bibinfo {author} {\bibfnamefont {J.~C.}\ \bibnamefont
				{Owrutsky}}, \bibinfo {author} {\bibfnamefont {D.}~\bibnamefont {Raftery}},\
			and\ \bibinfo {author} {\bibfnamefont {R.~M.}\ \bibnamefont {Hochstrasser}},\
		}\bibfield  {title} {\bibinfo {title} {{Vibrational Relaxation Dynamics in
					Solutions}},\ }\href {https://doi.org/10.1146/annurev.pc.45.100194.002511}
		{\bibfield  {journal} {\bibinfo  {journal} {Annu. Rev. Phys. Chem.}\ }\textbf
			{\bibinfo {volume} {45}},\ \bibinfo {pages} {519} (\bibinfo {year}
			{1994})}\BibitemShut {NoStop}%
		\bibitem [{\citenamefont {Delor}\ \emph {et~al.}(2014)\citenamefont {Delor},
			\citenamefont {Scattergood}, \citenamefont {Sazanovich}, \citenamefont
			{Parker}, \citenamefont {Greetham}, \citenamefont {Meijer}, \citenamefont
			{Towrie},\ and\ \citenamefont {Weinstein}}]{Delor2014Science}%
		\BibitemOpen
		\bibfield  {author} {\bibinfo {author} {\bibfnamefont {M.}~\bibnamefont
				{Delor}}, \bibinfo {author} {\bibfnamefont {P.~A.}\ \bibnamefont
				{Scattergood}}, \bibinfo {author} {\bibfnamefont {I.~V.}\ \bibnamefont
				{Sazanovich}}, \bibinfo {author} {\bibfnamefont {A.~W.}\ \bibnamefont
				{Parker}}, \bibinfo {author} {\bibfnamefont {G.~M.}\ \bibnamefont
				{Greetham}}, \bibinfo {author} {\bibfnamefont {A.~J. H.~M.}\ \bibnamefont
				{Meijer}}, \bibinfo {author} {\bibfnamefont {M.}~\bibnamefont {Towrie}},\
			and\ \bibinfo {author} {\bibfnamefont {J.~A.}\ \bibnamefont {Weinstein}},\
		}\bibfield  {title} {\bibinfo {title} {{Toward control of electron transfer
					in donor-acceptor molecules by bond-specific infrared excitation}},\ }\href
		{https://doi.org/10.1126/science.1259995} {\bibfield  {journal} {\bibinfo
				{journal} {Science}\ }\textbf {\bibinfo {volume} {346}},\ \bibinfo {pages}
			{1492} (\bibinfo {year} {2014})}\BibitemShut {NoStop}%
		\bibitem [{\citenamefont {Stensitzki}\ \emph {et~al.}(2018)\citenamefont
			{Stensitzki}, \citenamefont {Yang}, \citenamefont {Kozich}, \citenamefont
			{Ahmed}, \citenamefont {K{\"{o}}ssl}, \citenamefont {K{\"{u}}hn},\ and\
			\citenamefont {Heyne}}]{Stensitzki2018}%
		\BibitemOpen
		\bibfield  {author} {\bibinfo {author} {\bibfnamefont {T.}~\bibnamefont
				{Stensitzki}}, \bibinfo {author} {\bibfnamefont {Y.}~\bibnamefont {Yang}},
			\bibinfo {author} {\bibfnamefont {V.}~\bibnamefont {Kozich}}, \bibinfo
			{author} {\bibfnamefont {A.~A.}\ \bibnamefont {Ahmed}}, \bibinfo {author}
			{\bibfnamefont {F.}~\bibnamefont {K{\"{o}}ssl}}, \bibinfo {author}
			{\bibfnamefont {O.}~\bibnamefont {K{\"{u}}hn}},\ and\ \bibinfo {author}
			{\bibfnamefont {K.}~\bibnamefont {Heyne}},\ }\bibfield  {title} {\bibinfo
			{title} {{Acceleration of a ground-state reaction by selective
					femtosecond-infrared-laser-pulse excitation}},\ }\href
		{https://doi.org/10.1038/nchem.2909} {\bibfield  {journal} {\bibinfo
				{journal} {Nat. Chem.}\ }\textbf {\bibinfo {volume} {10}},\ \bibinfo {pages}
			{126} (\bibinfo {year} {2018})}\BibitemShut {NoStop}%
		\bibitem [{\citenamefont {Heyne}\ and\ \citenamefont
			{K{\"{u}}hn}(2019)}]{Heyne2019}%
		\BibitemOpen
		\bibfield  {author} {\bibinfo {author} {\bibfnamefont {K.}~\bibnamefont
				{Heyne}}\ and\ \bibinfo {author} {\bibfnamefont {O.}~\bibnamefont
				{K{\"{u}}hn}},\ }\bibfield  {title} {\bibinfo {title} {{Infrared Laser
					Excitation Controlled Reaction Acceleration in the Electronic Ground
					State}},\ }\href {https://doi.org/10.1021/jacs.9b02600} {\bibfield  {journal}
			{\bibinfo  {journal} {J. Am. Chem. Soc.}\ }\textbf {\bibinfo {volume}
				{141}},\ \bibinfo {pages} {11730} (\bibinfo {year} {2019})}\BibitemShut
		{NoStop}%
		\bibitem [{\citenamefont {Herrera}\ and\ \citenamefont
			{Owrutsky}(2020)}]{Herrera2019}%
		\BibitemOpen
		\bibfield  {author} {\bibinfo {author} {\bibfnamefont {F.}~\bibnamefont
				{Herrera}}\ and\ \bibinfo {author} {\bibfnamefont {J.}~\bibnamefont
				{Owrutsky}},\ }\bibfield  {title} {\bibinfo {title} {{Molecular polaritons
					for controlling chemistry with quantum optics}},\ }\href
		{https://doi.org/10.1063/1.5136320} {\bibfield  {journal} {\bibinfo
				{journal} {J. Chem. Phys.}\ }\textbf {\bibinfo {volume} {152}},\ \bibinfo
			{pages} {100902} (\bibinfo {year} {2020})}\BibitemShut {NoStop}%
		\bibitem [{\citenamefont {Xiang}\ and\ \citenamefont
			{Xiong}(2021)}]{Xiang2021JCP}%
		\BibitemOpen
		\bibfield  {author} {\bibinfo {author} {\bibfnamefont {B.}~\bibnamefont
				{Xiang}}\ and\ \bibinfo {author} {\bibfnamefont {W.}~\bibnamefont {Xiong}},\
		}\bibfield  {title} {\bibinfo {title} {{Molecular vibrational polariton: Its
					dynamics and potentials in novel chemistry and quantum technology}},\ }\href
		{https://doi.org/10.1063/5.0054896} {\bibfield  {journal} {\bibinfo
				{journal} {J. Chem. Phys.}\ }\textbf {\bibinfo {volume} {155}},\ \bibinfo
			{pages} {050901} (\bibinfo {year} {2021})}\BibitemShut {NoStop}%
		\bibitem [{\citenamefont {Garcia-Vidal}\ \emph {et~al.}(2021)\citenamefont
			{Garcia-Vidal}, \citenamefont {Ciuti},\ and\ \citenamefont
			{Ebbesen}}]{Garcia-Vidal2021}%
		\BibitemOpen
		\bibfield  {author} {\bibinfo {author} {\bibfnamefont {F.~J.}\ \bibnamefont
				{Garcia-Vidal}}, \bibinfo {author} {\bibfnamefont {C.}~\bibnamefont
				{Ciuti}},\ and\ \bibinfo {author} {\bibfnamefont {T.~W.}\ \bibnamefont
				{Ebbesen}},\ }\bibfield  {title} {\bibinfo {title} {{Manipulating Matter by
					Strong Coupling to Vacuum Fields}},\ }\href
		{https://doi.org/10.1126/science.abd0336} {\bibfield  {journal} {\bibinfo
				{journal} {Science}\ }\textbf {\bibinfo {volume} {373}},\ \bibinfo {pages}
			{eabd0336} (\bibinfo {year} {2021})}\BibitemShut {NoStop}%
		\bibitem [{\citenamefont {Li}\ \emph {et~al.}(2022)\citenamefont {Li},
			\citenamefont {Cui}, \citenamefont {Subotnik},\ and\ \citenamefont
			{Nitzan}}]{Li2022Review}%
		\BibitemOpen
		\bibfield  {author} {\bibinfo {author} {\bibfnamefont {T.~E.}\ \bibnamefont
				{Li}}, \bibinfo {author} {\bibfnamefont {B.}~\bibnamefont {Cui}}, \bibinfo
			{author} {\bibfnamefont {J.~E.}\ \bibnamefont {Subotnik}},\ and\ \bibinfo
			{author} {\bibfnamefont {A.}~\bibnamefont {Nitzan}},\ }\bibfield  {title}
		{\bibinfo {title} {{Molecular Polaritonics: Chemical Dynamics Under Strong
					Light–Matter Coupling}},\ }\bibfield  {journal} {\bibinfo  {journal} {Annu.
				Rev. Phys. Chem.}\ }\textbf {\bibinfo {volume} {73}},\ \href
		{https://doi.org/10.1146/annurev-physchem-090519-042621}
		{10.1146/annurev-physchem-090519-042621} (\bibinfo {year} {2022})\BibitemShut
		{NoStop}%
		\bibitem [{\citenamefont {Hutchison}\ \emph {et~al.}(2012)\citenamefont
			{Hutchison}, \citenamefont {Schwartz}, \citenamefont {Genet}, \citenamefont
			{Devaux},\ and\ \citenamefont {Ebbesen}}]{Hutchison2012}%
		\BibitemOpen
		\bibfield  {author} {\bibinfo {author} {\bibfnamefont {J.~A.}\ \bibnamefont
				{Hutchison}}, \bibinfo {author} {\bibfnamefont {T.}~\bibnamefont {Schwartz}},
			\bibinfo {author} {\bibfnamefont {C.}~\bibnamefont {Genet}}, \bibinfo
			{author} {\bibfnamefont {E.}~\bibnamefont {Devaux}},\ and\ \bibinfo {author}
			{\bibfnamefont {T.~W.}\ \bibnamefont {Ebbesen}},\ }\bibfield  {title}
		{\bibinfo {title} {{Modifying Chemical Landscapes by Coupling to Vacuum
					Fields}},\ }\href {https://doi.org/10.1002/ange.201107033} {\bibfield
			{journal} {\bibinfo  {journal} {Angew. Chemie}\ }\textbf {\bibinfo {volume}
				{124}},\ \bibinfo {pages} {1624} (\bibinfo {year} {2012})}\BibitemShut
		{NoStop}%
		\bibitem [{\citenamefont {Thomas}\ \emph {et~al.}(2016)\citenamefont {Thomas},
			\citenamefont {George}, \citenamefont {Shalabney}, \citenamefont {Dryzhakov},
			\citenamefont {Varma}, \citenamefont {Moran}, \citenamefont {Chervy},
			\citenamefont {Zhong}, \citenamefont {Devaux}, \citenamefont {Genet},
			\citenamefont {Hutchison},\ and\ \citenamefont {Ebbesen}}]{Thomas2016}%
		\BibitemOpen
		\bibfield  {author} {\bibinfo {author} {\bibfnamefont {A.}~\bibnamefont
				{Thomas}}, \bibinfo {author} {\bibfnamefont {J.}~\bibnamefont {George}},
			\bibinfo {author} {\bibfnamefont {A.}~\bibnamefont {Shalabney}}, \bibinfo
			{author} {\bibfnamefont {M.}~\bibnamefont {Dryzhakov}}, \bibinfo {author}
			{\bibfnamefont {S.~J.}\ \bibnamefont {Varma}}, \bibinfo {author}
			{\bibfnamefont {J.}~\bibnamefont {Moran}}, \bibinfo {author} {\bibfnamefont
				{T.}~\bibnamefont {Chervy}}, \bibinfo {author} {\bibfnamefont
				{X.}~\bibnamefont {Zhong}}, \bibinfo {author} {\bibfnamefont
				{E.}~\bibnamefont {Devaux}}, \bibinfo {author} {\bibfnamefont
				{C.}~\bibnamefont {Genet}}, \bibinfo {author} {\bibfnamefont {J.~A.}\
				\bibnamefont {Hutchison}},\ and\ \bibinfo {author} {\bibfnamefont {T.~W.}\
				\bibnamefont {Ebbesen}},\ }\bibfield  {title} {\bibinfo {title}
			{{Ground-State Chemical Reactivity under Vibrational Coupling to the Vacuum
					Electromagnetic Field}},\ }\href {https://doi.org/10.1002/anie.201605504}
		{\bibfield  {journal} {\bibinfo  {journal} {Angew. Chemie Int. Ed.}\ }\textbf
			{\bibinfo {volume} {55}},\ \bibinfo {pages} {11462} (\bibinfo {year}
			{2016})}\BibitemShut {NoStop}%
		\bibitem [{\citenamefont {Thomas}\ \emph {et~al.}(2019)\citenamefont {Thomas},
			\citenamefont {Lethuillier-Karl}, \citenamefont {Nagarajan}, \citenamefont
			{Vergauwe}, \citenamefont {George}, \citenamefont {Chervy}, \citenamefont
			{Shalabney}, \citenamefont {Devaux}, \citenamefont {Genet}, \citenamefont
			{Moran},\ and\ \citenamefont {Ebbesen}}]{Thomas2019_science}%
		\BibitemOpen
		\bibfield  {author} {\bibinfo {author} {\bibfnamefont {A.}~\bibnamefont
				{Thomas}}, \bibinfo {author} {\bibfnamefont {L.}~\bibnamefont
				{Lethuillier-Karl}}, \bibinfo {author} {\bibfnamefont {K.}~\bibnamefont
				{Nagarajan}}, \bibinfo {author} {\bibfnamefont {R.~M.~A.}\ \bibnamefont
				{Vergauwe}}, \bibinfo {author} {\bibfnamefont {J.}~\bibnamefont {George}},
			\bibinfo {author} {\bibfnamefont {T.}~\bibnamefont {Chervy}}, \bibinfo
			{author} {\bibfnamefont {A.}~\bibnamefont {Shalabney}}, \bibinfo {author}
			{\bibfnamefont {E.}~\bibnamefont {Devaux}}, \bibinfo {author} {\bibfnamefont
				{C.}~\bibnamefont {Genet}}, \bibinfo {author} {\bibfnamefont
				{J.}~\bibnamefont {Moran}},\ and\ \bibinfo {author} {\bibfnamefont {T.~W.}\
				\bibnamefont {Ebbesen}},\ }\bibfield  {title} {\bibinfo {title} {{Tilting a
					ground-state reactivity landscape by vibrational strong coupling}},\ }\href
		{https://doi.org/10.1126/science.aau7742} {\bibfield  {journal} {\bibinfo
				{journal} {Science}\ }\textbf {\bibinfo {volume} {363}},\ \bibinfo {pages}
			{615} (\bibinfo {year} {2019})}\BibitemShut {NoStop}%
		\bibitem [{\citenamefont {Imperatore}\ \emph {et~al.}(2021)\citenamefont
			{Imperatore}, \citenamefont {Asbury},\ and\ \citenamefont
			{Giebink}}]{Imperatore2021}%
		\BibitemOpen
		\bibfield  {author} {\bibinfo {author} {\bibfnamefont {M.~V.}\ \bibnamefont
				{Imperatore}}, \bibinfo {author} {\bibfnamefont {J.~B.}\ \bibnamefont
				{Asbury}},\ and\ \bibinfo {author} {\bibfnamefont {N.~C.}\ \bibnamefont
				{Giebink}},\ }\bibfield  {title} {\bibinfo {title} {{Reproducibility of
					cavity-enhanced chemical reaction rates in the vibrational strong coupling
					regime}},\ }\href {https://doi.org/10.1063/5.0046307} {\bibfield  {journal}
			{\bibinfo  {journal} {J. Chem. Phys.}\ }\textbf {\bibinfo {volume} {154}},\
			\bibinfo {pages} {191103} (\bibinfo {year} {2021})}\BibitemShut {NoStop}%
		\bibitem [{\citenamefont {Orgiu}\ \emph {et~al.}(2015)\citenamefont {Orgiu},
			\citenamefont {George}, \citenamefont {Hutchison}, \citenamefont {Devaux},
			\citenamefont {Dayen}, \citenamefont {Doudin}, \citenamefont {Stellacci},
			\citenamefont {Genet}, \citenamefont {Schachenmayer}, \citenamefont {Genes},
			\citenamefont {Pupillo}, \citenamefont {Samor{\`{i}}},\ and\ \citenamefont
			{Ebbesen}}]{Orgiu2015}%
		\BibitemOpen
		\bibfield  {author} {\bibinfo {author} {\bibfnamefont {E.}~\bibnamefont
				{Orgiu}}, \bibinfo {author} {\bibfnamefont {J.}~\bibnamefont {George}},
			\bibinfo {author} {\bibfnamefont {J.~A.}\ \bibnamefont {Hutchison}}, \bibinfo
			{author} {\bibfnamefont {E.}~\bibnamefont {Devaux}}, \bibinfo {author}
			{\bibfnamefont {J.~F.}\ \bibnamefont {Dayen}}, \bibinfo {author}
			{\bibfnamefont {B.}~\bibnamefont {Doudin}}, \bibinfo {author} {\bibfnamefont
				{F.}~\bibnamefont {Stellacci}}, \bibinfo {author} {\bibfnamefont
				{C.}~\bibnamefont {Genet}}, \bibinfo {author} {\bibfnamefont
				{J.}~\bibnamefont {Schachenmayer}}, \bibinfo {author} {\bibfnamefont
				{C.}~\bibnamefont {Genes}}, \bibinfo {author} {\bibfnamefont
				{G.}~\bibnamefont {Pupillo}}, \bibinfo {author} {\bibfnamefont
				{P.}~\bibnamefont {Samor{\`{i}}}},\ and\ \bibinfo {author} {\bibfnamefont
				{T.~W.}\ \bibnamefont {Ebbesen}},\ }\bibfield  {title} {\bibinfo {title}
			{{Conductivity in organic semiconductors hybridized with the vacuum field}},\
		}\href {https://doi.org/10.1038/nmat4392} {\bibfield  {journal} {\bibinfo
				{journal} {Nat. Mater.}\ }\textbf {\bibinfo {volume} {14}},\ \bibinfo {pages}
			{1123} (\bibinfo {year} {2015})}\BibitemShut {NoStop}%
		\bibitem [{\citenamefont {Coles}\ \emph {et~al.}(2014)\citenamefont {Coles},
			\citenamefont {Somaschi}, \citenamefont {Michetti}, \citenamefont {Clark},
			\citenamefont {Lagoudakis}, \citenamefont {Savvidis},\ and\ \citenamefont
			{Lidzey}}]{Coles2014}%
		\BibitemOpen
		\bibfield  {author} {\bibinfo {author} {\bibfnamefont {D.~M.}\ \bibnamefont
				{Coles}}, \bibinfo {author} {\bibfnamefont {N.}~\bibnamefont {Somaschi}},
			\bibinfo {author} {\bibfnamefont {P.}~\bibnamefont {Michetti}}, \bibinfo
			{author} {\bibfnamefont {C.}~\bibnamefont {Clark}}, \bibinfo {author}
			{\bibfnamefont {P.~G.}\ \bibnamefont {Lagoudakis}}, \bibinfo {author}
			{\bibfnamefont {P.~G.}\ \bibnamefont {Savvidis}},\ and\ \bibinfo {author}
			{\bibfnamefont {D.~G.}\ \bibnamefont {Lidzey}},\ }\bibfield  {title}
		{\bibinfo {title} {{Polariton-mediated energy transfer between organic dyes
					in a strongly coupled optical microcavity}},\ }\href
		{https://doi.org/10.1038/nmat3950} {\bibfield  {journal} {\bibinfo  {journal}
				{Nat. Mater.}\ }\textbf {\bibinfo {volume} {13}},\ \bibinfo {pages} {712}
			(\bibinfo {year} {2014})}\BibitemShut {NoStop}%
		\bibitem [{\citenamefont {Zhong}\ \emph {et~al.}(2017)\citenamefont {Zhong},
			\citenamefont {Chervy}, \citenamefont {Zhang}, \citenamefont {Thomas},
			\citenamefont {George}, \citenamefont {Genet}, \citenamefont {Hutchison},\
			and\ \citenamefont {Ebbesen}}]{Zhong2017}%
		\BibitemOpen
		\bibfield  {author} {\bibinfo {author} {\bibfnamefont {X.}~\bibnamefont
				{Zhong}}, \bibinfo {author} {\bibfnamefont {T.}~\bibnamefont {Chervy}},
			\bibinfo {author} {\bibfnamefont {L.}~\bibnamefont {Zhang}}, \bibinfo
			{author} {\bibfnamefont {A.}~\bibnamefont {Thomas}}, \bibinfo {author}
			{\bibfnamefont {J.}~\bibnamefont {George}}, \bibinfo {author} {\bibfnamefont
				{C.}~\bibnamefont {Genet}}, \bibinfo {author} {\bibfnamefont {J.~A.}\
				\bibnamefont {Hutchison}},\ and\ \bibinfo {author} {\bibfnamefont {T.~W.}\
				\bibnamefont {Ebbesen}},\ }\bibfield  {title} {\bibinfo {title} {{Energy
					Transfer between Spatially Separated Entangled Molecules}},\ }\href
		{https://doi.org/10.1002/anie.201703539} {\bibfield  {journal} {\bibinfo
				{journal} {Angew. Chemie Int. Ed.}\ }\textbf {\bibinfo {volume} {56}},\
			\bibinfo {pages} {9034} (\bibinfo {year} {2017})}\BibitemShut {NoStop}%
		\bibitem [{\citenamefont {S{\'{a}}ez-Bl{\'{a}}zquez}\ \emph
			{et~al.}(2018)\citenamefont {S{\'{a}}ez-Bl{\'{a}}zquez}, \citenamefont
			{Feist}, \citenamefont {Fern{\'{a}}ndez-Dom{\'{i}}nguez},\ and\ \citenamefont
			{Garc{\'{i}}a-Vidal}}]{Saez-Blazquez2018}%
		\BibitemOpen
		\bibfield  {author} {\bibinfo {author} {\bibfnamefont {R.}~\bibnamefont
				{S{\'{a}}ez-Bl{\'{a}}zquez}}, \bibinfo {author} {\bibfnamefont
				{J.}~\bibnamefont {Feist}}, \bibinfo {author} {\bibfnamefont {A.~I.}\
				\bibnamefont {Fern{\'{a}}ndez-Dom{\'{i}}nguez}},\ and\ \bibinfo {author}
			{\bibfnamefont {F.~J.}\ \bibnamefont {Garc{\'{i}}a-Vidal}},\ }\bibfield
		{title} {\bibinfo {title} {{Organic polaritons enable local vibrations to
					drive long-range energy transfer}},\ }\href
		{https://doi.org/10.1103/PhysRevB.97.241407} {\bibfield  {journal} {\bibinfo
				{journal} {Phys. Rev. B}\ }\textbf {\bibinfo {volume} {97}},\ \bibinfo
			{pages} {241407} (\bibinfo {year} {2018})}\BibitemShut {NoStop}%
		\bibitem [{\citenamefont {Du}\ \emph {et~al.}(2018)\citenamefont {Du},
			\citenamefont {Mart{\'{i}}nez-Mart{\'{i}}nez}, \citenamefont {Ribeiro},
			\citenamefont {Hu}, \citenamefont {Menon},\ and\ \citenamefont
			{Yuen-Zhou}}]{Du2018}%
		\BibitemOpen
		\bibfield  {author} {\bibinfo {author} {\bibfnamefont {M.}~\bibnamefont
				{Du}}, \bibinfo {author} {\bibfnamefont {L.~A.}\ \bibnamefont
				{Mart{\'{i}}nez-Mart{\'{i}}nez}}, \bibinfo {author} {\bibfnamefont {R.~F.}\
				\bibnamefont {Ribeiro}}, \bibinfo {author} {\bibfnamefont {Z.}~\bibnamefont
				{Hu}}, \bibinfo {author} {\bibfnamefont {V.~M.}\ \bibnamefont {Menon}},\ and\
			\bibinfo {author} {\bibfnamefont {J.}~\bibnamefont {Yuen-Zhou}},\ }\bibfield
		{title} {\bibinfo {title} {{Theory for polariton-assisted remote energy
					transfer}},\ }\href {https://doi.org/10.1039/C8SC00171E} {\bibfield
			{journal} {\bibinfo  {journal} {Chem. Sci.}\ }\textbf {\bibinfo {volume}
				{9}},\ \bibinfo {pages} {6659} (\bibinfo {year} {2018})}\BibitemShut
		{NoStop}%
		\bibitem [{\citenamefont {Georgiou}\ \emph {et~al.}(2021)\citenamefont
			{Georgiou}, \citenamefont {Jayaprakash}, \citenamefont {Othonos},\ and\
			\citenamefont {Lidzey}}]{Georgiou2021}%
		\BibitemOpen
		\bibfield  {author} {\bibinfo {author} {\bibfnamefont {K.}~\bibnamefont
				{Georgiou}}, \bibinfo {author} {\bibfnamefont {R.}~\bibnamefont
				{Jayaprakash}}, \bibinfo {author} {\bibfnamefont {A.}~\bibnamefont
				{Othonos}},\ and\ \bibinfo {author} {\bibfnamefont {D.~G.}\ \bibnamefont
				{Lidzey}},\ }\bibfield  {title} {\bibinfo {title} {{Ultralong‐Range
					Polariton‐Assisted Energy Transfer in Organic Microcavities}},\ }\href
		{https://doi.org/10.1002/anie.202105442} {\bibfield  {journal} {\bibinfo
				{journal} {Angew. Chemie Int. Ed.}\ }\textbf {\bibinfo {volume} {60}},\
			\bibinfo {pages} {16661} (\bibinfo {year} {2021})}\BibitemShut {NoStop}%
		\bibitem [{\citenamefont {Xiang}\ \emph {et~al.}(2020)\citenamefont {Xiang},
			\citenamefont {Ribeiro}, \citenamefont {Du}, \citenamefont {Chen},
			\citenamefont {Yang}, \citenamefont {Wang}, \citenamefont {Yuen-Zhou},\ and\
			\citenamefont {Xiong}}]{Xiang2020Science}%
		\BibitemOpen
		\bibfield  {author} {\bibinfo {author} {\bibfnamefont {B.}~\bibnamefont
				{Xiang}}, \bibinfo {author} {\bibfnamefont {R.~F.}\ \bibnamefont {Ribeiro}},
			\bibinfo {author} {\bibfnamefont {M.}~\bibnamefont {Du}}, \bibinfo {author}
			{\bibfnamefont {L.}~\bibnamefont {Chen}}, \bibinfo {author} {\bibfnamefont
				{Z.}~\bibnamefont {Yang}}, \bibinfo {author} {\bibfnamefont {J.}~\bibnamefont
				{Wang}}, \bibinfo {author} {\bibfnamefont {J.}~\bibnamefont {Yuen-Zhou}},\
			and\ \bibinfo {author} {\bibfnamefont {W.}~\bibnamefont {Xiong}},\ }\bibfield
		{title} {\bibinfo {title} {{Intermolecular vibrational energy transfer
					enabled by microcavity strong light–matter coupling}},\ }\href
		{https://doi.org/10.1126/science.aba3544} {\bibfield  {journal} {\bibinfo
				{journal} {Science}\ }\textbf {\bibinfo {volume} {368}},\ \bibinfo {pages}
			{665} (\bibinfo {year} {2020})}\BibitemShut {NoStop}%
		\bibitem [{\citenamefont {Shalabney}\ \emph {et~al.}(2015)\citenamefont
			{Shalabney}, \citenamefont {George}, \citenamefont {Hutchison}, \citenamefont
			{Pupillo}, \citenamefont {Genet},\ and\ \citenamefont
			{Ebbesen}}]{Shalabney2015}%
		\BibitemOpen
		\bibfield  {author} {\bibinfo {author} {\bibfnamefont {A.}~\bibnamefont
				{Shalabney}}, \bibinfo {author} {\bibfnamefont {J.}~\bibnamefont {George}},
			\bibinfo {author} {\bibfnamefont {J.}~\bibnamefont {Hutchison}}, \bibinfo
			{author} {\bibfnamefont {G.}~\bibnamefont {Pupillo}}, \bibinfo {author}
			{\bibfnamefont {C.}~\bibnamefont {Genet}},\ and\ \bibinfo {author}
			{\bibfnamefont {T.~W.}\ \bibnamefont {Ebbesen}},\ }\bibfield  {title}
		{\bibinfo {title} {{Coherent coupling of molecular resonators with a
					microcavity mode}},\ }\href {https://doi.org/10.1038/ncomms6981} {\bibfield
			{journal} {\bibinfo  {journal} {Nat. Commun.}\ }\textbf {\bibinfo {volume}
				{6}},\ \bibinfo {pages} {5981} (\bibinfo {year} {2015})}\BibitemShut
		{NoStop}%
		\bibitem [{\citenamefont {Long}\ and\ \citenamefont
			{Simpkins}(2015)}]{Long2015}%
		\BibitemOpen
		\bibfield  {author} {\bibinfo {author} {\bibfnamefont {J.~P.}\ \bibnamefont
				{Long}}\ and\ \bibinfo {author} {\bibfnamefont {B.~S.}\ \bibnamefont
				{Simpkins}},\ }\bibfield  {title} {\bibinfo {title} {{Coherent Coupling
					between a Molecular Vibration and Fabry–Perot Optical Cavity to Give
					Hybridized States in the Strong Coupling Limit}},\ }\href
		{https://doi.org/10.1021/ph5003347} {\bibfield  {journal} {\bibinfo
				{journal} {ACS Photonics}\ }\textbf {\bibinfo {volume} {2}},\ \bibinfo
			{pages} {130} (\bibinfo {year} {2015})}\BibitemShut {NoStop}%
		\bibitem [{\citenamefont {Xiang}\ \emph {et~al.}(2019)\citenamefont {Xiang},
			\citenamefont {Ribeiro}, \citenamefont {Chen}, \citenamefont {Wang},
			\citenamefont {Du}, \citenamefont {Yuen-Zhou},\ and\ \citenamefont
			{Xiong}}]{Xiang2019State}%
		\BibitemOpen
		\bibfield  {author} {\bibinfo {author} {\bibfnamefont {B.}~\bibnamefont
				{Xiang}}, \bibinfo {author} {\bibfnamefont {R.~F.}\ \bibnamefont {Ribeiro}},
			\bibinfo {author} {\bibfnamefont {L.}~\bibnamefont {Chen}}, \bibinfo {author}
			{\bibfnamefont {J.}~\bibnamefont {Wang}}, \bibinfo {author} {\bibfnamefont
				{M.}~\bibnamefont {Du}}, \bibinfo {author} {\bibfnamefont {J.}~\bibnamefont
				{Yuen-Zhou}},\ and\ \bibinfo {author} {\bibfnamefont {W.}~\bibnamefont
				{Xiong}},\ }\bibfield  {title} {\bibinfo {title} {{State-Selective Polariton
					to Dark State Relaxation Dynamics}},\ }\href
		{https://doi.org/10.1021/acs.jpca.9b04601} {\bibfield  {journal} {\bibinfo
				{journal} {J. Phys. Chem. A}\ }\textbf {\bibinfo {volume} {123}},\ \bibinfo
			{pages} {5918} (\bibinfo {year} {2019})}\BibitemShut {NoStop}%
		\bibitem [{\citenamefont {Li}\ \emph {et~al.}(2021{\natexlab{a}})\citenamefont
			{Li}, \citenamefont {Nitzan},\ and\ \citenamefont
			{Subotnik}}]{Li2020Nonlinear}%
		\BibitemOpen
		\bibfield  {author} {\bibinfo {author} {\bibfnamefont {T.~E.}\ \bibnamefont
				{Li}}, \bibinfo {author} {\bibfnamefont {A.}~\bibnamefont {Nitzan}},\ and\
			\bibinfo {author} {\bibfnamefont {J.~E.}\ \bibnamefont {Subotnik}},\
		}\bibfield  {title} {\bibinfo {title} {{Cavity molecular dynamics simulations
					of vibrational polariton-enhanced molecular nonlinear absorption}},\ }\href
		{https://doi.org/10.1063/5.0037623} {\bibfield  {journal} {\bibinfo
				{journal} {J. Chem. Phys.}\ }\textbf {\bibinfo {volume} {154}},\ \bibinfo
			{pages} {094124} (\bibinfo {year} {2021}{\natexlab{a}})}\BibitemShut
		{NoStop}%
		\bibitem [{\citenamefont {Ribeiro}\ \emph {et~al.}(2020)\citenamefont
			{Ribeiro}, \citenamefont {Campos-Gonzalez-Angulo}, \citenamefont {Giebink},
			\citenamefont {Xiong},\ and\ \citenamefont {Yuen-Zhou}}]{Ribeiro2020}%
		\BibitemOpen
		\bibfield  {author} {\bibinfo {author} {\bibfnamefont {R.~F.}\ \bibnamefont
				{Ribeiro}}, \bibinfo {author} {\bibfnamefont {J.~A.}\ \bibnamefont
				{Campos-Gonzalez-Angulo}}, \bibinfo {author} {\bibfnamefont {N.~C.}\
				\bibnamefont {Giebink}}, \bibinfo {author} {\bibfnamefont {W.}~\bibnamefont
				{Xiong}},\ and\ \bibinfo {author} {\bibfnamefont {J.}~\bibnamefont
				{Yuen-Zhou}},\ }\bibfield  {title} {\bibinfo {title} {{Enhanced optical
					nonlinearities under strong light-matter coupling}},\ }\href
		{http://arxiv.org/abs/2006.08519} {\  (\bibinfo {year} {2020})},\ \Eprint
		{https://arxiv.org/abs/2006.08519} {arXiv:2006.08519} \BibitemShut {NoStop}%
		\bibitem [{\citenamefont {Campos-Gonzalez-Angulo}\ and\ \citenamefont
			{Yuen-Zhou}(2020)}]{Campos-Gonzalez-Angulo2020}%
		\BibitemOpen
		\bibfield  {author} {\bibinfo {author} {\bibfnamefont {J.~A.}\ \bibnamefont
				{Campos-Gonzalez-Angulo}}\ and\ \bibinfo {author} {\bibfnamefont
				{J.}~\bibnamefont {Yuen-Zhou}},\ }\bibfield  {title} {\bibinfo {title}
			{{Polaritonic normal modes in transition state theory}},\ }\href
		{https://doi.org/10.1063/5.0007547} {\bibfield  {journal} {\bibinfo
				{journal} {J. Chem. Phys.}\ }\textbf {\bibinfo {volume} {152}},\ \bibinfo
			{pages} {161101} (\bibinfo {year} {2020})}\BibitemShut {NoStop}%
		\bibitem [{\citenamefont {Li}\ \emph {et~al.}(2021{\natexlab{b}})\citenamefont
			{Li}, \citenamefont {Mandal},\ and\ \citenamefont {Huo}}]{LiHuo2020}%
		\BibitemOpen
		\bibfield  {author} {\bibinfo {author} {\bibfnamefont {X.}~\bibnamefont
				{Li}}, \bibinfo {author} {\bibfnamefont {A.}~\bibnamefont {Mandal}},\ and\
			\bibinfo {author} {\bibfnamefont {P.}~\bibnamefont {Huo}},\ }\bibfield
		{title} {\bibinfo {title} {{Cavity frequency-dependent theory for vibrational
					polariton chemistry}},\ }\href {https://doi.org/10.1038/s41467-021-21610-9}
		{\bibfield  {journal} {\bibinfo  {journal} {Nat. Commun.}\ }\textbf {\bibinfo
				{volume} {12}},\ \bibinfo {pages} {1315} (\bibinfo {year}
			{2021}{\natexlab{b}})}\BibitemShut {NoStop}%
		\bibitem [{\citenamefont {Li}\ \emph {et~al.}(2020)\citenamefont {Li},
			\citenamefont {Subotnik},\ and\ \citenamefont {Nitzan}}]{Li2020Water}%
		\BibitemOpen
		\bibfield  {author} {\bibinfo {author} {\bibfnamefont {T.~E.}\ \bibnamefont
				{Li}}, \bibinfo {author} {\bibfnamefont {J.~E.}\ \bibnamefont {Subotnik}},\
			and\ \bibinfo {author} {\bibfnamefont {A.}~\bibnamefont {Nitzan}},\
		}\bibfield  {title} {\bibinfo {title} {{Cavity molecular dynamics simulations
					of liquid water under vibrational ultrastrong coupling}},\ }\href
		{https://doi.org/10.1073/pnas.2009272117} {\bibfield  {journal} {\bibinfo
				{journal} {Proc. Natl. Acad. Sci.}\ }\textbf {\bibinfo {volume} {117}},\
			\bibinfo {pages} {18324} (\bibinfo {year} {2020})}\BibitemShut {NoStop}%
		\bibitem [{\citenamefont {Sidler}\ \emph {et~al.}(2021)\citenamefont {Sidler},
			\citenamefont {Sch{\"{a}}fer}, \citenamefont {Ruggenthaler},\ and\
			\citenamefont {Rubio}}]{Sidler2021}%
		\BibitemOpen
		\bibfield  {author} {\bibinfo {author} {\bibfnamefont {D.}~\bibnamefont
				{Sidler}}, \bibinfo {author} {\bibfnamefont {C.}~\bibnamefont
				{Sch{\"{a}}fer}}, \bibinfo {author} {\bibfnamefont {M.}~\bibnamefont
				{Ruggenthaler}},\ and\ \bibinfo {author} {\bibfnamefont {A.}~\bibnamefont
				{Rubio}},\ }\bibfield  {title} {\bibinfo {title} {{Polaritonic Chemistry:
					Collective Strong Coupling Implies Strong Local Modification of Chemical
					Properties}},\ }\href {https://doi.org/10.1021/acs.jpclett.0c03436}
		{\bibfield  {journal} {\bibinfo  {journal} {J. Phys. Chem. Lett.}\ }\textbf
			{\bibinfo {volume} {12}},\ \bibinfo {pages} {508} (\bibinfo {year}
			{2021})}\BibitemShut {NoStop}%
		\bibitem [{\citenamefont {Flick}\ \emph {et~al.}(2017)\citenamefont {Flick},
			\citenamefont {Ruggenthaler}, \citenamefont {Appel},\ and\ \citenamefont
			{Rubio}}]{Flick2017}%
		\BibitemOpen
		\bibfield  {author} {\bibinfo {author} {\bibfnamefont {J.}~\bibnamefont
				{Flick}}, \bibinfo {author} {\bibfnamefont {M.}~\bibnamefont {Ruggenthaler}},
			\bibinfo {author} {\bibfnamefont {H.}~\bibnamefont {Appel}},\ and\ \bibinfo
			{author} {\bibfnamefont {A.}~\bibnamefont {Rubio}},\ }\bibfield  {title}
		{\bibinfo {title} {{Atoms and Molecules in Cavities, from Weak to Strong
					Coupling in Quantum-Electrodynamics (QED) Chemistry}},\ }\href
		{https://doi.org/10.1073/pnas.1615509114} {\bibfield  {journal} {\bibinfo
				{journal} {Proc. Natl. Acad. Sci.}\ }\textbf {\bibinfo {volume} {114}},\
			\bibinfo {pages} {3026} (\bibinfo {year} {2017})}\BibitemShut {NoStop}%
		\bibitem [{\citenamefont {Haugland}\ \emph {et~al.}(2020)\citenamefont
			{Haugland}, \citenamefont {Ronca}, \citenamefont {Kj{\o}nstad}, \citenamefont
			{Rubio},\ and\ \citenamefont {Koch}}]{Haugland2020}%
		\BibitemOpen
		\bibfield  {author} {\bibinfo {author} {\bibfnamefont {T.~S.}\ \bibnamefont
				{Haugland}}, \bibinfo {author} {\bibfnamefont {E.}~\bibnamefont {Ronca}},
			\bibinfo {author} {\bibfnamefont {E.~F.}\ \bibnamefont {Kj{\o}nstad}},
			\bibinfo {author} {\bibfnamefont {A.}~\bibnamefont {Rubio}},\ and\ \bibinfo
			{author} {\bibfnamefont {H.}~\bibnamefont {Koch}},\ }\bibfield  {title}
		{\bibinfo {title} {{Coupled Cluster Theory for Molecular Polaritons: Changing
					Ground and Excited States}},\ }\href
		{https://doi.org/10.1103/PhysRevX.10.041043} {\bibfield  {journal} {\bibinfo
				{journal} {Phys. Rev. X}\ }\textbf {\bibinfo {volume} {10}},\ \bibinfo
			{pages} {041043} (\bibinfo {year} {2020})}\BibitemShut {NoStop}%
		\bibitem [{\citenamefont {Triana}\ \emph {et~al.}(2020)\citenamefont {Triana},
			\citenamefont {Hern{\'{a}}ndez},\ and\ \citenamefont
			{Herrera}}]{Triana2020Shape}%
		\BibitemOpen
		\bibfield  {author} {\bibinfo {author} {\bibfnamefont {J.~F.}\ \bibnamefont
				{Triana}}, \bibinfo {author} {\bibfnamefont {F.~J.}\ \bibnamefont
				{Hern{\'{a}}ndez}},\ and\ \bibinfo {author} {\bibfnamefont {F.}~\bibnamefont
				{Herrera}},\ }\bibfield  {title} {\bibinfo {title} {{The shape of the
					electric dipole function determines the sub-picosecond dynamics of anharmonic
					vibrational polaritons}},\ }\href {https://doi.org/10.1063/5.0009869}
		{\bibfield  {journal} {\bibinfo  {journal} {J. Chem. Phys.}\ }\textbf
			{\bibinfo {volume} {152}},\ \bibinfo {pages} {234111} (\bibinfo {year}
			{2020})},\ \Eprint {https://arxiv.org/abs/2003.07783} {arXiv:2003.07783}
		\BibitemShut {NoStop}%
		\bibitem [{\citenamefont {Luk}\ \emph {et~al.}(2017)\citenamefont {Luk},
			\citenamefont {Feist}, \citenamefont {Toppari},\ and\ \citenamefont
			{Groenhof}}]{Luk2017}%
		\BibitemOpen
		\bibfield  {author} {\bibinfo {author} {\bibfnamefont {H.~L.}\ \bibnamefont
				{Luk}}, \bibinfo {author} {\bibfnamefont {J.}~\bibnamefont {Feist}}, \bibinfo
			{author} {\bibfnamefont {J.~J.}\ \bibnamefont {Toppari}},\ and\ \bibinfo
			{author} {\bibfnamefont {G.}~\bibnamefont {Groenhof}},\ }\bibfield  {title}
		{\bibinfo {title} {{Multiscale Molecular Dynamics Simulations of Polaritonic
					Chemistry}},\ }\href {https://doi.org/10.1021/acs.jctc.7b00388} {\bibfield
			{journal} {\bibinfo  {journal} {J. Chem. Theory Comput.}\ }\textbf {\bibinfo
				{volume} {13}},\ \bibinfo {pages} {4324} (\bibinfo {year}
			{2017})}\BibitemShut {NoStop}%
		\bibitem [{\citenamefont {Li}\ \emph {et~al.}(2021{\natexlab{c}})\citenamefont
			{Li}, \citenamefont {Nitzan},\ and\ \citenamefont
			{Subotnik}}]{Li2021Collective}%
		\BibitemOpen
		\bibfield  {author} {\bibinfo {author} {\bibfnamefont {T.~E.}\ \bibnamefont
				{Li}}, \bibinfo {author} {\bibfnamefont {A.}~\bibnamefont {Nitzan}},\ and\
			\bibinfo {author} {\bibfnamefont {J.~E.}\ \bibnamefont {Subotnik}},\
		}\bibfield  {title} {\bibinfo {title} {{Collective vibrational strong
					coupling effects on molecular vibrational relaxation and energy transfer:
					Numerical insights via cavity molecular dynamics simulations}},\ }\href
		{https://doi.org/10.1002/anie.202103920} {\bibfield  {journal} {\bibinfo
				{journal} {Angew. Chemie Int. Ed.}\ ,\ \bibinfo {pages} {anie.202103920}}
			(\bibinfo {year} {2021}{\natexlab{c}})}\BibitemShut {NoStop}%
		\bibitem [{\citenamefont {Gaigeot}\ and\ \citenamefont
			{Sprik}(2003)}]{Gaigeot2003}%
		\BibitemOpen
		\bibfield  {author} {\bibinfo {author} {\bibfnamefont {M.-P.}\ \bibnamefont
				{Gaigeot}}\ and\ \bibinfo {author} {\bibfnamefont {M.}~\bibnamefont
				{Sprik}},\ }\bibfield  {title} {\bibinfo {title} {{Ab Initio Molecular
					Dynamics Computation of the Infrared Spectrum of Aqueous Uracil}},\ }\href
		{https://doi.org/10.1021/jp034788u} {\bibfield  {journal} {\bibinfo
				{journal} {J. Phys. Chem. B}\ }\textbf {\bibinfo {volume} {107}},\ \bibinfo
			{pages} {10344} (\bibinfo {year} {2003})}\BibitemShut {NoStop}%
		\bibitem [{\citenamefont {Brawley}\ \emph {et~al.}(2021)\citenamefont
			{Brawley}, \citenamefont {Storm}, \citenamefont {{Contreras Mora}},
			\citenamefont {Pelton},\ and\ \citenamefont {Sheldon}}]{Brawley2021}%
		\BibitemOpen
		\bibfield  {author} {\bibinfo {author} {\bibfnamefont {Z.~T.}\ \bibnamefont
				{Brawley}}, \bibinfo {author} {\bibfnamefont {S.~D.}\ \bibnamefont {Storm}},
			\bibinfo {author} {\bibfnamefont {D.~A.}\ \bibnamefont {{Contreras Mora}}},
			\bibinfo {author} {\bibfnamefont {M.}~\bibnamefont {Pelton}},\ and\ \bibinfo
			{author} {\bibfnamefont {M.}~\bibnamefont {Sheldon}},\ }\bibfield  {title}
		{\bibinfo {title} {{Angle-independent plasmonic substrates for multi-mode
					vibrational strong coupling with molecular thin films}},\ }\href
		{https://doi.org/10.1063/5.0039195} {\bibfield  {journal} {\bibinfo
				{journal} {J. Chem. Phys.}\ }\textbf {\bibinfo {volume} {154}},\ \bibinfo
			{pages} {104305} (\bibinfo {year} {2021})}\BibitemShut {NoStop}%
		\bibitem [{\citenamefont {Yoo}\ \emph {et~al.}(2021)\citenamefont {Yoo},
			\citenamefont {de~Le{\'{o}}n-P{\'{e}}rez}, \citenamefont {Pelton},
			\citenamefont {Lee}, \citenamefont {Mohr}, \citenamefont {Raschke},
			\citenamefont {Caldwell}, \citenamefont {Mart{\'{i}}n-Moreno},\ and\
			\citenamefont {Oh}}]{Yoo2021}%
		\BibitemOpen
		\bibfield  {author} {\bibinfo {author} {\bibfnamefont {D.}~\bibnamefont
				{Yoo}}, \bibinfo {author} {\bibfnamefont {F.}~\bibnamefont
				{de~Le{\'{o}}n-P{\'{e}}rez}}, \bibinfo {author} {\bibfnamefont
				{M.}~\bibnamefont {Pelton}}, \bibinfo {author} {\bibfnamefont {I.-H.}\
				\bibnamefont {Lee}}, \bibinfo {author} {\bibfnamefont {D.~A.}\ \bibnamefont
				{Mohr}}, \bibinfo {author} {\bibfnamefont {M.~B.}\ \bibnamefont {Raschke}},
			\bibinfo {author} {\bibfnamefont {J.~D.}\ \bibnamefont {Caldwell}}, \bibinfo
			{author} {\bibfnamefont {L.}~\bibnamefont {Mart{\'{i}}n-Moreno}},\ and\
			\bibinfo {author} {\bibfnamefont {S.-H.}\ \bibnamefont {Oh}},\ }\bibfield
		{title} {\bibinfo {title} {{Ultrastrong plasmon–phonon coupling via
					epsilon-near-zero nanocavities}},\ }\href
		{https://doi.org/10.1038/s41566-020-00731-5} {\bibfield  {journal} {\bibinfo
				{journal} {Nat. Photonics}\ }\textbf {\bibinfo {volume} {15}},\ \bibinfo
			{pages} {125} (\bibinfo {year} {2021})}\BibitemShut {NoStop}%
		\bibitem [{\citenamefont {Grafton}\ \emph {et~al.}(2021)\citenamefont
			{Grafton}, \citenamefont {Dunkelberger}, \citenamefont {Simpkins},
			\citenamefont {Triana}, \citenamefont {Hern{\'{a}}ndez}, \citenamefont
			{Herrera},\ and\ \citenamefont {Owrutsky}}]{Grafton2020}%
		\BibitemOpen
		\bibfield  {author} {\bibinfo {author} {\bibfnamefont {A.~B.}\ \bibnamefont
				{Grafton}}, \bibinfo {author} {\bibfnamefont {A.~D.}\ \bibnamefont
				{Dunkelberger}}, \bibinfo {author} {\bibfnamefont {B.~S.}\ \bibnamefont
				{Simpkins}}, \bibinfo {author} {\bibfnamefont {J.~F.}\ \bibnamefont
				{Triana}}, \bibinfo {author} {\bibfnamefont {F.~J.}\ \bibnamefont
				{Hern{\'{a}}ndez}}, \bibinfo {author} {\bibfnamefont {F.}~\bibnamefont
				{Herrera}},\ and\ \bibinfo {author} {\bibfnamefont {J.~C.}\ \bibnamefont
				{Owrutsky}},\ }\bibfield  {title} {\bibinfo {title} {{Excited-state
					vibration-polariton transitions and dynamics in nitroprusside}},\ }\href
		{https://doi.org/10.1038/s41467-020-20535-z} {\bibfield  {journal} {\bibinfo
				{journal} {Nat. Commun.}\ }\textbf {\bibinfo {volume} {12}},\ \bibinfo
			{pages} {214} (\bibinfo {year} {2021})}\BibitemShut {NoStop}%
		\bibitem [{\citenamefont {Sch{\"{a}}fer}\ \emph {et~al.}(2020)\citenamefont
			{Sch{\"{a}}fer}, \citenamefont {Ruggenthaler}, \citenamefont {Rokaj},\ and\
			\citenamefont {Rubio}}]{Schafer2020}%
		\BibitemOpen
		\bibfield  {author} {\bibinfo {author} {\bibfnamefont {C.}~\bibnamefont
				{Sch{\"{a}}fer}}, \bibinfo {author} {\bibfnamefont {M.}~\bibnamefont
				{Ruggenthaler}}, \bibinfo {author} {\bibfnamefont {V.}~\bibnamefont
				{Rokaj}},\ and\ \bibinfo {author} {\bibfnamefont {A.}~\bibnamefont {Rubio}},\
		}\bibfield  {title} {\bibinfo {title} {{Relevance of the Quadratic
					Diamagnetic and Self-Polarization Terms in Cavity Quantum Electrodynamics}},\
		}\href {https://doi.org/10.1021/acsphotonics.9b01649} {\bibfield  {journal}
			{\bibinfo  {journal} {ACS Photonics}\ }\textbf {\bibinfo {volume} {7}},\
			\bibinfo {pages} {975} (\bibinfo {year} {2020})}\BibitemShut {NoStop}%
		\bibitem [{\citenamefont {Taylor}\ \emph {et~al.}(2020)\citenamefont {Taylor},
			\citenamefont {Mandal}, \citenamefont {Zhou},\ and\ \citenamefont
			{Huo}}]{Taylor2020}%
		\BibitemOpen
		\bibfield  {author} {\bibinfo {author} {\bibfnamefont {M.~A.~D.}\
				\bibnamefont {Taylor}}, \bibinfo {author} {\bibfnamefont {A.}~\bibnamefont
				{Mandal}}, \bibinfo {author} {\bibfnamefont {W.}~\bibnamefont {Zhou}},\ and\
			\bibinfo {author} {\bibfnamefont {P.}~\bibnamefont {Huo}},\ }\bibfield
		{title} {\bibinfo {title} {{Resolution of Gauge Ambiguities in Molecular
					Cavity Quantum Electrodynamics}},\ }\href
		{https://doi.org/10.1103/PhysRevLett.125.123602} {\bibfield  {journal}
			{\bibinfo  {journal} {Phys. Rev. Lett.}\ }\textbf {\bibinfo {volume} {125}},\
			\bibinfo {pages} {123602} (\bibinfo {year} {2020})}\BibitemShut {NoStop}%
		\bibitem [{\citenamefont {Stokes}\ and\ \citenamefont
			{Nazir}(2020)}]{Stokes2020}%
		\BibitemOpen
		\bibfield  {author} {\bibinfo {author} {\bibfnamefont {A.}~\bibnamefont
				{Stokes}}\ and\ \bibinfo {author} {\bibfnamefont {A.}~\bibnamefont {Nazir}},\
		}\bibfield  {title} {\bibinfo {title} {{Implications of gauge-freedom for
					nonrelativistic quantum electrodynamics}},\ }\href
		{http://arxiv.org/abs/2009.10662} {\  (\bibinfo {year} {2020})},\ \Eprint
		{https://arxiv.org/abs/2009.10662} {arXiv:2009.10662} \BibitemShut {NoStop}%
		\bibitem [{\citenamefont {Heidelbach}\ \emph {et~al.}(1998)\citenamefont
			{Heidelbach}, \citenamefont {Fedchenia}, \citenamefont {Schwarzer},\ and\
			\citenamefont {Schroeder}}]{Heidelbach1998}%
		\BibitemOpen
		\bibfield  {author} {\bibinfo {author} {\bibfnamefont {C.}~\bibnamefont
				{Heidelbach}}, \bibinfo {author} {\bibfnamefont {I.~I.}\ \bibnamefont
				{Fedchenia}}, \bibinfo {author} {\bibfnamefont {D.}~\bibnamefont
				{Schwarzer}},\ and\ \bibinfo {author} {\bibfnamefont {J.}~\bibnamefont
				{Schroeder}},\ }\bibfield  {title} {\bibinfo {title} {{Molecular-dynamics
					simulation of collisional energy transfer from vibrationally highly excited
					azulene in compressed CO2}},\ }\href {https://doi.org/10.1063/1.476474}
		{\bibfield  {journal} {\bibinfo  {journal} {J. Chem. Phys.}\ }\textbf
			{\bibinfo {volume} {108}},\ \bibinfo {pages} {10152} (\bibinfo {year}
			{1998})}\BibitemShut {NoStop}%
		\bibitem [{\citenamefont {Kabadi}\ and\ \citenamefont
			{Rice}(2004)}]{Kabadi2004}%
		\BibitemOpen
		\bibfield  {author} {\bibinfo {author} {\bibfnamefont {V.~N.}\ \bibnamefont
				{Kabadi}}\ and\ \bibinfo {author} {\bibfnamefont {B.~M.}\ \bibnamefont
				{Rice}},\ }\bibfield  {title} {\bibinfo {title} {{Molecular Dynamics
					Simulations of Normal Mode Vibrational Energy Transfer in Liquid
					Nitromethane}},\ }\href {https://doi.org/10.1021/jp035975v} {\bibfield
			{journal} {\bibinfo  {journal} {J. Phys. Chem. A}\ }\textbf {\bibinfo
				{volume} {108}},\ \bibinfo {pages} {532} (\bibinfo {year}
			{2004})}\BibitemShut {NoStop}%
		\bibitem [{\citenamefont {Kandratsenka}\ \emph {et~al.}(2009)\citenamefont
			{Kandratsenka}, \citenamefont {Schroeder}, \citenamefont {Schwarzer},\ and\
			\citenamefont {Vikhrenko}}]{Kandratsenka2009}%
		\BibitemOpen
		\bibfield  {author} {\bibinfo {author} {\bibfnamefont {A.}~\bibnamefont
				{Kandratsenka}}, \bibinfo {author} {\bibfnamefont {J.}~\bibnamefont
				{Schroeder}}, \bibinfo {author} {\bibfnamefont {D.}~\bibnamefont
				{Schwarzer}},\ and\ \bibinfo {author} {\bibfnamefont {V.~S.}\ \bibnamefont
				{Vikhrenko}},\ }\bibfield  {title} {\bibinfo {title} {{Nonequilibrium
					molecular dynamics simulations of vibrational energy relaxation of HOD in
					D2O}},\ }\href {https://doi.org/10.1063/1.3126781} {\bibfield  {journal}
			{\bibinfo  {journal} {J. Chem. Phys.}\ }\textbf {\bibinfo {volume} {130}},\
			\bibinfo {pages} {174507} (\bibinfo {year} {2009})}\BibitemShut {NoStop}%
		\bibitem [{\citenamefont {Miller}(1978)}]{Miller1978}%
		\BibitemOpen
		\bibfield  {author} {\bibinfo {author} {\bibfnamefont {W.~H.}\ \bibnamefont
				{Miller}},\ }\bibfield  {title} {\bibinfo {title} {{A Classical/Semiclassical
					Theory for the Interaction of Infrared Radiation with Molecular Systems}},\
		}\href {https://doi.org/10.1063/1.436793} {\bibfield  {journal} {\bibinfo
				{journal} {J. Chem. Phys.}\ }\textbf {\bibinfo {volume} {69}},\ \bibinfo
			{pages} {2188} (\bibinfo {year} {1978})}\BibitemShut {NoStop}%
		\bibitem [{\citenamefont {Orel}\ and\ \citenamefont {Miller}(1980)}]{Orel1980}%
		\BibitemOpen
		\bibfield  {author} {\bibinfo {author} {\bibfnamefont {A.~E.}\ \bibnamefont
				{Orel}}\ and\ \bibinfo {author} {\bibfnamefont {W.~H.}\ \bibnamefont
				{Miller}},\ }\bibfield  {title} {\bibinfo {title} {{Classical model for
					laser‐induced nonadiabatic collision processes}},\ }\href
		{https://doi.org/10.1063/1.439923} {\bibfield  {journal} {\bibinfo  {journal}
				{J. Chem. Phys.}\ }\textbf {\bibinfo {volume} {73}},\ \bibinfo {pages} {241}
			(\bibinfo {year} {1980})}\BibitemShut {NoStop}%
		\bibitem [{\citenamefont {Vergauwe}\ \emph {et~al.}(2019)\citenamefont
			{Vergauwe}, \citenamefont {Thomas}, \citenamefont {Nagarajan}, \citenamefont
			{Shalabney}, \citenamefont {George}, \citenamefont {Chervy}, \citenamefont
			{Seidel}, \citenamefont {Devaux}, \citenamefont {Torbeev},\ and\
			\citenamefont {Ebbesen}}]{Vergauwe2019}%
		\BibitemOpen
		\bibfield  {author} {\bibinfo {author} {\bibfnamefont {R.~M.~A.}\
				\bibnamefont {Vergauwe}}, \bibinfo {author} {\bibfnamefont {A.}~\bibnamefont
				{Thomas}}, \bibinfo {author} {\bibfnamefont {K.}~\bibnamefont {Nagarajan}},
			\bibinfo {author} {\bibfnamefont {A.}~\bibnamefont {Shalabney}}, \bibinfo
			{author} {\bibfnamefont {J.}~\bibnamefont {George}}, \bibinfo {author}
			{\bibfnamefont {T.}~\bibnamefont {Chervy}}, \bibinfo {author} {\bibfnamefont
				{M.}~\bibnamefont {Seidel}}, \bibinfo {author} {\bibfnamefont
				{E.}~\bibnamefont {Devaux}}, \bibinfo {author} {\bibfnamefont
				{V.}~\bibnamefont {Torbeev}},\ and\ \bibinfo {author} {\bibfnamefont {T.~W.}\
				\bibnamefont {Ebbesen}},\ }\bibfield  {title} {\bibinfo {title}
			{{Modification of Enzyme Activity by Vibrational Strong Coupling of Water}},\
		}\href {https://doi.org/10.1002/anie.201908876} {\bibfield  {journal}
			{\bibinfo  {journal} {Angew. Chemie Int. Ed.}\ }\textbf {\bibinfo {volume}
				{58}},\ \bibinfo {pages} {15324} (\bibinfo {year} {2019})}\BibitemShut
		{NoStop}%
		\bibitem [{\citenamefont {Xiang}\ \emph {et~al.}(2018)\citenamefont {Xiang},
			\citenamefont {Ribeiro}, \citenamefont {Dunkelberger}, \citenamefont {Wang},
			\citenamefont {Li}, \citenamefont {Simpkins}, \citenamefont {Owrutsky},
			\citenamefont {Yuen-Zhou},\ and\ \citenamefont {Xiong}}]{Xiang2018}%
		\BibitemOpen
		\bibfield  {author} {\bibinfo {author} {\bibfnamefont {B.}~\bibnamefont
				{Xiang}}, \bibinfo {author} {\bibfnamefont {R.~F.}\ \bibnamefont {Ribeiro}},
			\bibinfo {author} {\bibfnamefont {A.~D.}\ \bibnamefont {Dunkelberger}},
			\bibinfo {author} {\bibfnamefont {J.}~\bibnamefont {Wang}}, \bibinfo {author}
			{\bibfnamefont {Y.}~\bibnamefont {Li}}, \bibinfo {author} {\bibfnamefont
				{B.~S.}\ \bibnamefont {Simpkins}}, \bibinfo {author} {\bibfnamefont {J.~C.}\
				\bibnamefont {Owrutsky}}, \bibinfo {author} {\bibfnamefont {J.}~\bibnamefont
				{Yuen-Zhou}},\ and\ \bibinfo {author} {\bibfnamefont {W.}~\bibnamefont
				{Xiong}},\ }\bibfield  {title} {\bibinfo {title} {{Two-dimensional infrared
					spectroscopy of vibrational polaritons}},\ }\href
		{https://doi.org/10.1073/pnas.1722063115} {\bibfield  {journal} {\bibinfo
				{journal} {Proc. Natl. Acad. Sci.}\ }\textbf {\bibinfo {volume} {115}},\
			\bibinfo {pages} {4845} (\bibinfo {year} {2018})}\BibitemShut {NoStop}%
		\bibitem [{\citenamefont {Haugland}\ \emph {et~al.}(2021)\citenamefont
			{Haugland}, \citenamefont {Sch{\"{a}}fer}, \citenamefont {Ronca},
			\citenamefont {Rubio},\ and\ \citenamefont {Koch}}]{Haugland2021}%
		\BibitemOpen
		\bibfield  {author} {\bibinfo {author} {\bibfnamefont {T.~S.}\ \bibnamefont
				{Haugland}}, \bibinfo {author} {\bibfnamefont {C.}~\bibnamefont
				{Sch{\"{a}}fer}}, \bibinfo {author} {\bibfnamefont {E.}~\bibnamefont
				{Ronca}}, \bibinfo {author} {\bibfnamefont {A.}~\bibnamefont {Rubio}},\ and\
			\bibinfo {author} {\bibfnamefont {H.}~\bibnamefont {Koch}},\ }\bibfield
		{title} {\bibinfo {title} {{Intermolecular interactions in optical cavities:
					An ab initio QED study}},\ }\href {https://doi.org/10.1063/5.0039256}
		{\bibfield  {journal} {\bibinfo  {journal} {J. Chem. Phys.}\ }\textbf
			{\bibinfo {volume} {154}},\ \bibinfo {pages} {094113} (\bibinfo {year}
			{2021})},\ \Eprint {https://arxiv.org/abs/2012.01080} {arXiv:2012.01080}
		\BibitemShut {NoStop}%
		\bibitem [{\citenamefont {Kapil}\ \emph {et~al.}(2019)\citenamefont {Kapil},
			\citenamefont {Rossi}, \citenamefont {Marsalek}, \citenamefont {Petraglia},
			\citenamefont {Litman}, \citenamefont {Spura}, \citenamefont {Cheng},
			\citenamefont {Cuzzocrea}, \citenamefont {Mei{\ss}ner}, \citenamefont
			{Wilkins}, \citenamefont {Helfrecht}, \citenamefont {Juda}, \citenamefont
			{Bienvenue}, \citenamefont {Fang}, \citenamefont {Kessler}, \citenamefont
			{Poltavsky}, \citenamefont {Vandenbrande}, \citenamefont {Wieme},
			\citenamefont {Corminboeuf}, \citenamefont {K{\"{u}}hne}, \citenamefont
			{Manolopoulos}, \citenamefont {Markland}, \citenamefont {Richardson},
			\citenamefont {Tkatchenko}, \citenamefont {Tribello}, \citenamefont {{Van
					Speybroeck}},\ and\ \citenamefont {Ceriotti}}]{Kapil2019}%
		\BibitemOpen
		\bibfield  {author} {\bibinfo {author} {\bibfnamefont {V.}~\bibnamefont
				{Kapil}}, \bibinfo {author} {\bibfnamefont {M.}~\bibnamefont {Rossi}},
			\bibinfo {author} {\bibfnamefont {O.}~\bibnamefont {Marsalek}}, \bibinfo
			{author} {\bibfnamefont {R.}~\bibnamefont {Petraglia}}, \bibinfo {author}
			{\bibfnamefont {Y.}~\bibnamefont {Litman}}, \bibinfo {author} {\bibfnamefont
				{T.}~\bibnamefont {Spura}}, \bibinfo {author} {\bibfnamefont
				{B.}~\bibnamefont {Cheng}}, \bibinfo {author} {\bibfnamefont
				{A.}~\bibnamefont {Cuzzocrea}}, \bibinfo {author} {\bibfnamefont {R.~H.}\
				\bibnamefont {Mei{\ss}ner}}, \bibinfo {author} {\bibfnamefont {D.~M.}\
				\bibnamefont {Wilkins}}, \bibinfo {author} {\bibfnamefont {B.~A.}\
				\bibnamefont {Helfrecht}}, \bibinfo {author} {\bibfnamefont {P.}~\bibnamefont
				{Juda}}, \bibinfo {author} {\bibfnamefont {S.~P.}\ \bibnamefont {Bienvenue}},
			\bibinfo {author} {\bibfnamefont {W.}~\bibnamefont {Fang}}, \bibinfo {author}
			{\bibfnamefont {J.}~\bibnamefont {Kessler}}, \bibinfo {author} {\bibfnamefont
				{I.}~\bibnamefont {Poltavsky}}, \bibinfo {author} {\bibfnamefont
				{S.}~\bibnamefont {Vandenbrande}}, \bibinfo {author} {\bibfnamefont
				{J.}~\bibnamefont {Wieme}}, \bibinfo {author} {\bibfnamefont
				{C.}~\bibnamefont {Corminboeuf}}, \bibinfo {author} {\bibfnamefont {T.~D.}\
				\bibnamefont {K{\"{u}}hne}}, \bibinfo {author} {\bibfnamefont {D.~E.}\
				\bibnamefont {Manolopoulos}}, \bibinfo {author} {\bibfnamefont {T.~E.}\
				\bibnamefont {Markland}}, \bibinfo {author} {\bibfnamefont {J.~O.}\
				\bibnamefont {Richardson}}, \bibinfo {author} {\bibfnamefont
				{A.}~\bibnamefont {Tkatchenko}}, \bibinfo {author} {\bibfnamefont {G.~A.}\
				\bibnamefont {Tribello}}, \bibinfo {author} {\bibfnamefont {V.}~\bibnamefont
				{{Van Speybroeck}}},\ and\ \bibinfo {author} {\bibfnamefont {M.}~\bibnamefont
				{Ceriotti}},\ }\bibfield  {title} {\bibinfo {title} {{i-PI 2.0: A universal
					force engine for advanced molecular simulations}},\ }\href
		{https://doi.org/10.1016/j.cpc.2018.09.020} {\bibfield  {journal} {\bibinfo
				{journal} {Comput. Phys. Commun.}\ }\textbf {\bibinfo {volume} {236}},\
			\bibinfo {pages} {214} (\bibinfo {year} {2019})}\BibitemShut {NoStop}%
		\bibitem [{\citenamefont {Plimpton}(1995)}]{Plimpton1995}%
		\BibitemOpen
		\bibfield  {author} {\bibinfo {author} {\bibfnamefont {S.}~\bibnamefont
				{Plimpton}},\ }\bibfield  {title} {\bibinfo {title} {{Fast Parallel
					Algorithms for Short-Range Molecular Dynamics}},\ }\href
		{https://doi.org/10.1006/jcph.1995.1039} {\bibfield  {journal} {\bibinfo
				{journal} {J. Comput. Phys.}\ }\textbf {\bibinfo {volume} {117}},\ \bibinfo
			{pages} {1} (\bibinfo {year} {1995})}\BibitemShut {NoStop}%
		\bibitem [{\citenamefont {Li}(2022)}]{TELi2020Github}%
		\BibitemOpen
		\bibfield  {author} {\bibinfo {author} {\bibfnamefont {T.~E.}\ \bibnamefont
				{Li}},\ }\bibfield  {title} {\bibinfo {title} {{Cavity Molecular Dynamics
					Simulations Tool Sets (v0.1)}},\ }\href@noop {} {\bibfield  {journal}
			{\bibinfo  {journal} {Zenodo}\ ,\ \bibinfo {pages}
				{https://doi.org/10.5281/zenodo.6629538}} (\bibinfo {year}
			{2022})}\BibitemShut {NoStop}%
	\end{thebibliography}
	
	%

		\section*{Acknowledgments}
	This material is based upon work supported by the U.S. National Science Foundation under Grant No. CHE1953701 (A.N.); 
	and US Department of Energy, Office of Science,
	Basic Energy Sciences, Chemical Sciences, Geosciences, and Biosciences Division under Award No. DE-SC0019397  (J.E.S.).
	
	\section*{Author Contributions}
	T.E.L. conceived the project, performed simulations and plotted the figures, T.E.L., A.N., and J.E.S. discussed the results and wrote the paper.
	
	\section*{Competing interests}
	The authors declare no competing interests.
	
\onecolumngrid

\clearpage
\thispagestyle{plain}

\pagenumbering{gobble}



\section*{Supplementary material}

## SUPPLEMENTARY METHODS

Following [1, 2], given a full-quantum photon-electron-nuclei Hamiltonian in the dipole gauge and under long-wave approximation, cavity molecular dynamics (CavMD) simulations invoke the Born–Oppenheimer approximation. The full quantum Hamiltonian is projected onto the electronic ground state, leading to:

$$\hat{H}_{\mathrm{QED}}^{\mathrm{G}} = \hat{H}_{\mathrm{M}}^{\mathrm{G}} + \hat{H}_{\mathrm{F}}^{\mathrm{G}}. \tag{S1a}$$

Here, $\hat{H}_{\mathrm{M}}^{\mathrm{G}}$ denotes the standard ground-state molecular Hamiltonian, which is composed of kinetic energies of nuclei ($\hat{\mathbf{P}}_{nj}^2/2M_{nj}$), intramolecular interactions ($\hat{V}_g^{(n)}$) and pairwise intermolecular interactions ($\hat{V}_{\mathrm{inter}}^{(nl)}$):

$$\hat{H}_{\mathrm{M}}^{\mathrm{G}} = \sum_{n=1}^{N} \left( \sum_{j \in n} \frac{\hat{\mathbf{P}}_{nj}^2}{2M_{nj}} + \hat{V}_g^{(n)}(\{\hat{\mathbf{R}}_{nj}\}) \right) + \sum_{n=1}^{N} \sum_{l>n} \hat{V}_{\mathrm{inter}}^{(nl)}, \tag{S1b}$$

where $\hat{\mathbf{P}}_{nj}$, $\hat{\mathbf{R}}_{nj}$, and $M_{nj}$ denote the momentum operator, position operator, and mass for the $j$-th nucleus in molecule $n$. $\hat{H}_{\mathrm{F}}^{\mathrm{G}}$ denotes the field-related Hamiltonian [1, 2] where each cavity photon mode with wave vector $\mathbf{k}$ and polarization direction $\boldsymbol{\xi}_\lambda$ interacts with the total dipole moment of the molecular subsystem:

$$\hat{H}_{\mathrm{F}}^{\mathrm{G}} = \sum_{k,\lambda} \frac{\hat{\tilde{p}}_{k,\lambda}^2}{2m_{k,\lambda}} + \frac{1}{2} m_{k,\lambda} \omega_{k,\lambda}^2 \left( \hat{\tilde{q}}_{k,\lambda} + \sum_{n=1}^{N} \frac{\hat{d}_{ng,\lambda}}{\omega_{k,\lambda} \sqrt{\Omega \epsilon_0 m_{k,\lambda}}} \right)^2. \tag{S1c}$$

Here, $\hat{\tilde{p}}_{k,\lambda}$, $\hat{\tilde{q}}_{k,\lambda}$, $\omega_{k,\lambda}$, and $m_{k,\lambda}$ denote the momentum operator, position operator, frequency, and the auxiliary mass for each cavity photon. $\Omega$ denotes the volume for the cavity, $\epsilon_0$ denotes the vacuum permittivity, and $\hat{d}_{ng,\lambda}$ denotes the electronic ground-state dipole operator for molecule $n$ ($\hat{\mathbf{d}}_n$) projected along the direction of $\boldsymbol{\xi}_\lambda$: $\hat{d}_{ng,\lambda} \equiv \hat{\mathbf{d}}_n \cdot \boldsymbol{\xi}_\lambda$. Compared with usual light-matter Hamiltonian [3–6] where there is no auxiliary mass $m_{k,\lambda}$ term for the photonic operators ($\hat{p}_{k,\lambda}$ and $\hat{q}_{k,\lambda}$), in Eq. (S1c) $m_{k,\lambda}$ is introduced because standard molecular dynamics packages require this quantity. One can obtain $m_{k,\lambda}$ by simply rescaling the conventional photonic operators as follows: $\hat{q}_{k,\lambda} \equiv \sqrt{m_{k,\lambda}} \hat{\tilde{q}}_{k,\lambda}$ and $\hat{p}_{k,\lambda} \equiv \hat{\tilde{p}}_{k,\lambda}/\sqrt{m_{k,\lambda}}$. Hence, the use of the auxiliary mass does not alter any dynamics. In simulations, we take $m_{k,\lambda} = 1$ a.u. (atomic units) for simplicity.

---

* taoli@sas.upenn.edu
† anitzan@sas.upenn.edu
‡ subotnik@sas.upenn.edu

In Eq. (S1c), we can further define the coupling strength between each cavity photon and each individual molecule as

$$\varepsilon_{k,\lambda} \equiv \sqrt{m_{k,\lambda}\omega_{k,\lambda}^2/\Omega\epsilon_0}. \tag{S2}$$

In order to efficiently simulate the above quantum Hamiltonian in Eq. (S1), we reduce all of the operators to classical observables and obtain the equations of motion as follows

$$M_{nj}\ddot{\mathbf{R}}_{nj} = \mathbf{F}_{nj}^{(0)} - \sum_{k,\lambda}\left(\varepsilon_{k,\lambda}\widetilde{q}_{k,\lambda} + \frac{\varepsilon_{k,\lambda}^2}{m_{k,\lambda}\omega_{k,\lambda}^2}\sum_{l=1}^{N}d_{lg,\lambda}\right)\frac{\partial d_{ng,\lambda}}{\partial \mathbf{R}_{nj}}, \tag{S3a}$$

$$m_{k,\lambda}\ddot{\widetilde{q}}_{k,\lambda} = -m_{k,\lambda}\omega_{k,\lambda}^2\widetilde{q}_{k,\lambda} - \varepsilon_{k,\lambda}\sum_{n=1}^{N}d_{ng,\lambda}. \tag{S3b}$$

Here, $\mathbf{F}_{nj}^{(0)} = -\partial V_g^{(n)}/\partial \mathbf{R}_{nj} - \sum_{l\neq n}\partial V_{\text{inter}}^{(nl)}/\partial \mathbf{R}_{nj}$ denotes the molecular part of the force on each nuclei, i.e., the nuclear force outside the cavity.

a. *Periodic boundary conditions* VSC inside a Fabry–Pérot microcavity usually involves a macroscopic number (say, $10^9 \sim 10^{11}$) of molecules. For such a large molecular system, it is very expensive if we propagate (S3) directly. Hence, we further apply periodic boundary conditions, i.e., we assume that the molecular subsystem can be divided into $N_{\text{cell}}$ identical periodic cells. In detail, for Eq. (S3) we approximate the total dipole moment of molecules as $\sum_{n=1}^{N}d_{ng,\lambda} = N_{\text{cell}}\sum_{n=1}^{N_{\text{sub}}}d_{ng,\lambda}$, where $N_{\text{sub}} = N/N_{\text{cell}}$ denotes the number of molecules in a single cell. By further denoting $\widetilde{\widetilde{q}}_{k,\lambda} = \widetilde{q}_{k,\lambda}/\sqrt{N_{\text{cell}}}$, $\widetilde{\varepsilon}_{k,\lambda} = \sqrt{N_{\text{cell}}}\varepsilon_{k,\lambda}$, we can rewrite the equations of motion in (S3) in a symmetric form

$$M_{nj}\ddot{\mathbf{R}}_{nj} = \mathbf{F}_{nj}^{(0)} + \mathbf{F}_{nj}^{\text{cav}}, \tag{S4a}$$

$$m_{k,\lambda}\ddot{\widetilde{\widetilde{q}}}_{k,\lambda} = -m_{k,\lambda}\omega_{k,\lambda}^2\widetilde{\widetilde{q}}_{k,\lambda} - \widetilde{\varepsilon}_{k,\lambda}\sum_{n=1}^{N_{\text{sub}}}d_{ng,\lambda}. \tag{S4b}$$

Here,

$$\mathbf{F}_{nj}^{\text{cav}} = -\sum_{k,\lambda}(\widetilde{\varepsilon}_{k,\lambda}\widetilde{\widetilde{q}}_{k,\lambda} + \frac{\widetilde{\varepsilon}_{k,\lambda}^2}{m_{k,\lambda}\omega_{k,\lambda}^2}\sum_{l=1}^{N_{\text{sub}}}d_{lg,\lambda})\frac{\partial d_{ng,\lambda}}{\partial \mathbf{R}_{nj}} \tag{S4c}$$

denotes the cavity force on each nucleus. Because we invoke periodic boundary conditions, it is logical to redefine

$$\widetilde{\varepsilon}_{k,\lambda} = \sqrt{N_{\text{cell}}}\varepsilon_{k,\lambda}, \tag{S4d}$$

to characterize the effective coupling between the cavity mode and each molecule in the periodic cell [where $\varepsilon_{k,\lambda}$, the true coupling strength between each cavity photon and individual

3

molecules, has been defined in Eq. (S2)]. Note that the equations of motion under periodic boundary conditions (Eq. (S4)) have a similar form as those without the periodic boundary conditions (Eq. (S3)). Not surprisingly, if we take $N_{\text{cell}} = 1$ (or if we assume $\widetilde{\varepsilon}_{k,\lambda} = \varepsilon_{k,\lambda}$), Eq. (S4) reduces to Eq. (S3).

b. *Interpreting simulation results as Fabry–Pérot microcavities* If CavMD is used to simulate VSC phenomena in Fabry–Pérot microcavities with $N$ molecules and an experimental Rabi splitting $\Omega_N$, since CavMD simulations use periodic boundary conditions with $N_{\text{sub}}$ molecules in a periodic simulation cell, we can choose $\widetilde{\varepsilon}_{k,\lambda} \propto \sqrt{N_{\text{cell}}}$ so as to fit the experimentally observed Rabi splitting $\Omega_N$.

Such a choice of $\widetilde{\varepsilon}_{k,\lambda}$, however, overestimates the light-matter coupling strength between the cavity modes and each individual molecule by a factor of $\sqrt{N_{\text{cell}}}$ relative to the true coupling strength $\varepsilon_{k,\lambda}$; see Eq. (S4d). Hence, when a VSC effect on molecular properties is obtained via CavMD, it is necessary to check the dependence on periodic boundary conditions (or the choice of the system size $N_{\text{sub}}$) to make sure that any observed VSC effect is not an artifact of the simulation.

In the main text, we have chosen to check the system size (or periodic boundary conditions) dependence as follows [2, 7]: (i) we keep the molecular density the same; (ii) when $N_{\text{sub}}$ is enlarged by a factor of two, the effective light-matter coupling strength $\widetilde{\varepsilon}_{k,\lambda}$ is rescaled by a factor of $1/\sqrt{2}$ — such a balance between molecular number and coupling strength can maintain a fixed Rabi splitting; (iii) finally, by plotting the simulated VSC effect versus $N_{\text{sub}}$, the asymptotic results when $N_{\text{sub}}$ approaches a macroscopic number should correspond to Fabry–Pérot microcavities.

This procedure for enlarging the system size has an experimental analog. As shown in Supplementary Figure 1, experimentally one can imagine enlarging the cavity length by a factor of two, which increases the cavity volume by a factor of two. Such an enlarging process can ensure both cavities have a cavity mode with the same frequency (red waves), although the light-matter coupling in the larger volume cavity is just $1/\sqrt{2}$ of the smaller one. When molecules with the same density are placed inside the cavities, the change of the light-matter coupling and the molecular number (which is proportional to the cavity volume) balance each other, yielding an unchanged Rabi splitting. This experimental approach for investigating large cavities matches our CavMD approach exactly.

c. *Interpreting simulation results as plasmonic nanocavities* Of course, CavMD can also be used to simulate plasmonic nanocavities under VSC where the molecular number

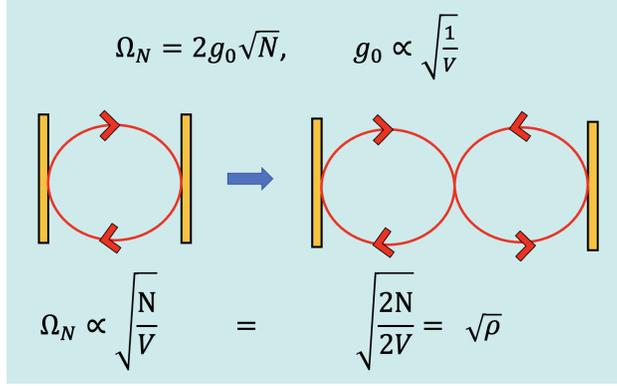

Supplementary Figure 1. Cartoon for the cavity size enlarging procedure. The Rabi splitting is the same when the cavity volume is increased by a factor of two and when molecules with the same density are filled in the cavity.

can be relatively small. Here, in CavMD we take $N_{\text{cell}} = 1$, so that $\widetilde{\varepsilon}_{k,\lambda} = \varepsilon_{k,\lambda}$ becomes the true coupling strength. Hence, for CavMD simulations with a given $N_{\text{sub}}$, one can interpret our CavMD dynamics as reporting on cavities with $N_{\text{sub}}$ molecules form VSC. However, since plasmonic cavities usually have very large cavity loss, in order to obtain a convincing result for simulating plasmonic nanocavities, we need to add a very large cavity loss on top of CavMD.

    *d. Simulating external electric field driving* When using CavMD to simulate driving nonequilibrium dynamics with an external electric field, we encounter a non-obvious choice as to what subsystem observable (internal to the cavity) we will choose to directly interact with the external electric field — e.g., will the external field drive directly the molecular vibrations or the electromagnetic cavity mode? In practice, we find that for nonequilibirum polariton dynamics, this choice of subsystem observable is immaterial: exciting either the (photonic or molecular) subsystem effectively pumps the polariton due to the fast coherent energy exchange between the two subsystems.

    Mathematically, on top of the equations of motion in Eq. (S4), we can introduce the external driving on either the cavity modes or the molecules as follows. Assuming the interaction between the external field and the coupled cavity-molecular system as

$$V_{\text{ext}}(t) = -\boldsymbol{\mu}_S \cdot \mathbf{E}_{\text{ext}}(t) - \sum_{k,\lambda} \mu_c^{k,\lambda} \boldsymbol{\xi}_\lambda \cdot \mathbf{E}_{\text{ext}}(t) \quad (S5)$$

where $\boldsymbol{\mu}_S$ denotes the total dipole moment vector of the molecular system, $\mu_c^{k,\lambda} \equiv Q_c^{k,\lambda} \widetilde{\widetilde{q}}_{k,\lambda}$ denotes the effective transition dipole moment (and $Q_c^{k,\lambda}$ denotes the effective charge) of the

cavity mode, and $\mathbf{E}_{\text{ext}}(t)$ denotes the external time-dependent driving field, we can rewrite the equations of motion of the coupled cavity-molecular system as follows:

$$M_{nj}\ddot{\mathbf{R}}_{nj} = \mathbf{F}_{nj}^{(0)} + \mathbf{F}_{nj}^{\text{cav}} + \mathbf{F}_{nj}^{\text{ext}}(t), \tag{S6a}$$

$$m_{k,\lambda}\ddot{\widetilde{\widetilde{q}}}_{k,\lambda} = -m_{k,\lambda}\omega_{k,\lambda}^2 \widetilde{\widetilde{q}}_{k,\lambda} - \widetilde{\varepsilon}_{k,\lambda}\sum_{n=1}^{N_{\text{sub}}} d_{ng,\lambda} + \mathbf{F}_{k,\lambda}^{\text{ext}}(t). \tag{S6b}$$

Here, the new added terms — the external force on each nucleus ($\mathbf{F}_{nj}^{\text{ext}}(t)$) or cavity mode ($\mathbf{F}_{k,\lambda}^{\text{ext}}(t)$) is defined as:

$$\mathbf{F}_{nj}^{\text{ext}}(t) = Q_{nj}\mathbf{E}_{\text{ext}}(t), \tag{S7a}$$

$$\mathbf{F}_{k,\lambda}^{\text{ext}}(t) = Q_c^{k,\lambda}\mathbf{E}_{\text{ext}}(t), \tag{S7b}$$

where $Q_{nj}$ denotes the partial charge of each nucleus (if a classical force field is used to simulate the molecular subsystem).

In the main text, we have chosen to drive the molecular subsystem only (by setting $\mu_c = Q_c = 0$) because this choice avoids introducing an additional parameter $Q_c$, the value of which varies for different cavities and could hinder the universality of our finding. In Supplementary Figure 5, we will show that if the driving of the molecular subsystem is turned off and when the external field pumps only on the cavity modes (with a finite $Q_c$), a consistent simulation result is obtained as compared with case of pure molecular driving (where $Q_c = 0$; which is reported in the main text).

e. *Simulation details* We prepare 216 $^{12}$CO$_2$ molecules in a cubic cell with cell length 24.292 Å (45.905 a.u.) [8], which leads to the density of the liquid $^{12}$CO$_2$ system as 1.101 g/cm$^3$. We assume that a cavity formed by a pair of two parallel mirrors is placed along the $z$-axis. For a liquid $^{12}$CO$_2$ system under VSC, we consider only one cavity mode (but with two polarization directions $x$ and $y$) at 2320 cm$^{-1}$ (the peak frequency of $^{12}$C=O asymmetric stretch), which is coupled to the molecular subsystem with an effective coupling strength $\widetilde{\varepsilon} = 2 \times 10^{-4}$ a.u. or $4 \times 10^{-4}$ a.u.

Before pumping the system with an external laser pulse, we perform simulations to equilibrate the system. At 300 K, we first run the simulation for 150 ps to guarantee thermal equilibrium under a NVT (constant particle number, volume, and temperature) ensemble where a Langevin thermostat with a lifetime (i.e., inverse friction) of 100 fs is applied to the momenta of all particles (nuclei + photons). The resulting equilibrium configurations

are used as starting points for 40 consecutive NVE (constant particle number, volume, and energy) trajectories of length 20 ps. At the beginning of each trajectory the velocities are resampled by a Maxwell-Boltzmann distribution under 300 K. The intermolecular Coulombic interactions are calculated by an Ewald summation. The simulation step is set as 0.5 fs and we store the snapshots of trajectories every 2 fs.

We next perform nonequilibrium simulations under an external pulse. We start each simulation with an equilibrium geometry, which is chosen from the starting configurations of the above 40 NVE trajectories. At the beginning stage of each NVE trajectory, we apply an external pulse

$$\mathbf{E}_{\text{ext}}(t) = E_0 \cos(\omega t + \phi) \mathbf{e}_x \tag{S8}$$

to excite the molecular subsystem only. Here, the phase $\phi \in [0, 2\pi)$ is set as random. This pulse is turned on at $t_{\text{start}} = 0.1$ ps and is turned off at $t_{\text{end}} = 0.6$ ps. The incoming field is strong: $E_0 = 3.084 \times 10^7$ V/m ($6 \times 10^{-3}$ a.u.) and the fluence of the pulse is $F = 632$ mJ/cm$^2$. Note that in Supplementary Figure 5 we perform CavMD simulations by assuming that the incoming field is coupled to the cavity mode only; in Supplementary Figure 6, we also perform simulations when the incoming field is replaced by a Gaussian pulse.

Each nonequilibrium trajectory is run for 20 ps under a NVE ensemble and the physical properties are calculated by averaging over these 40 nonequilibrium trajectories. Note that the use of NVE trajectories when calculating equilibrium or nonequilibrium physical properties implies the assumption of zero cavity loss.

Apart from these NVE nonequilibrium simulations, in Supplementary Figure 4 below we also attach a Langevin thermostat with a friction lifetime $\tau_c$ to the photonic momenta to propagate the nonequilibrium simulations. Such a treatment can mimic the situation of a non-zero cavity loss, where the cavity lifetime is $\tau_c$.

For the simulation of the liquid mixture of $^{13}$CO$_2$ and $^{12}$CO$_2$, we simply replace some $^{12}$C atoms by the $^{13}$C atoms and rerun the above equilibrium + nonequilibrium simulations.

As stated in the manuscript and above, we refer to our cavity system as "a single cavity mode" because the photonic energies for the two polarization directions are the same, and only a single photon peak appears in the IR spectroscopy — so there is only a single mode *from a spectroscopic point of view*. Of course, we could also refer to our cavity system as a system with "two energy-degenerated cavity photons with different polarization directions", but this description seems a bit lengthy.

7

*f. Definition of the $CO_2$ force field*  The $CO_2$ force field used in the manuscript [2] largely resembles Ref. [9] except for the use of an anharmonic C=O bond potential instead of a harmonic one. According to Ref. [9], in the harmonic limit, this force field can capture macroscopic thermal equilibrium observables and also the equilibrium IR spectroscopy of liquid $CO_2$ that are consistent with experiments.

In this $CO_2$ force field, the intramolecular interaction is characterized by

$$V_{\mathrm{g}}^{(n)} = V_{\mathrm{CO}}(R_{n1}) + V_{\mathrm{CO}}(R_{n2}) + \frac{1}{2}k_\theta \left(\theta_n - \theta_{\mathrm{eq}}\right)^2. \tag{S9}$$

Here, $R_{n1}$ and $R_{n2}$ denote the lengths of two C=O bonds, and $\theta_n$ and $\theta_{\mathrm{eq}}$ denote the O−C−O angle and the equilibrium angle. The form of $V_{\mathrm{CO}}$ is defined as follows:

$$V_{\mathrm{CO}}(r) = D_r \left[\alpha_r^2 \Delta r^2 - \alpha_r^3 \Delta r^3 + \frac{7}{12}\alpha_r^4 \Delta r^4\right], \tag{S10}$$

where $\Delta r = r - r_{\mathrm{eq}}$. Eq. (S10) is a fourth-order Taylor expansion of a Morse potential $V_{\mathrm{M}}(r) = D_r[1 - \exp(-\alpha_r \Delta r)]^2$. The parameters are taken as follows: $r_{\mathrm{eq}} = 1.162$ Å (2.196 a.u.), $D_r = 127.13$ kcal mol$^{-1}$ (0.2026 a.u.), and $\alpha_r = 2.819$ Å$^{-1}$ (1.492 a.u.) are chosen to fit the harmonic potential used in Ref. [9] in the harmonic limit and the value of $D_r$ takes the bond dissociation energy of O=CO at room temperature [10]. The other parameters in Eq. (S9) are given as $k_\theta = 108.0$ kJ mol$^{-1}$rad$^{-2}$ (0.0861 a.u.), and $\theta_{\mathrm{eq}} = 180$ deg.

The pairwise intermolecular potential is characterized by the Lennard–Jones potential ($V_{\mathrm{LJ}}^{(nl)}$) plus the Coulombic interactions between atoms ($V_{\mathrm{Coul}}^{(nl)}$):

$$V_{\mathrm{inter}}^{(nl)} = V_{\mathrm{LJ}}^{(nl)} + V_{\mathrm{Coul}}^{(nl)}, \tag{S11}$$

where the superscript $n, l$ denote the indices of different molecules. The form of $V_{\mathrm{LJ}}^{(nl)}$ is

$$V_{\mathrm{LJ}}^{(nl)} = \sum_{i \in n} \sum_{j \in l} 4\epsilon_{ij} \left[\left(\frac{\sigma_{ij}}{R_{ij}}\right)^{12} - \left(\frac{\sigma_{ij}}{R_{ij}}\right)^6\right], \tag{S12}$$

where $R_{ij}$ ($i \in n$ and $j \in l$) denotes the distance between atoms in molecules $n$ and $l$. For parameters, $\epsilon_{\mathrm{CC}} = 0.0559$ kcal mol$^{-1}$ ($8.9126 \times 10^{-5}$ a.u.), $\sigma_{\mathrm{CC}} = 2.800$ Å (5.291 a.u.), $\epsilon_{\mathrm{OO}} = 0.1597$ kcal mol$^{-1}$ ($2.5454 \times 10^{-4}$ a.u.), $\sigma_{\mathrm{OO}} = 3.028$ Å (5.722 a.u.), $\epsilon_{\mathrm{CO}} = \sqrt{\epsilon_{\mathrm{CC}}\epsilon_{\mathrm{OO}}}$, and $\sigma_{\mathrm{CO}} = (\sigma_{\mathrm{CC}} + \sigma_{\mathrm{OO}})/2$. The form of $V_{\mathrm{Coul}}^{(nl)}$ is

$$V_{\mathrm{Coul}}^{(nl)} = \sum_{i \in n} \sum_{j \in l} \frac{Q_i Q_j}{4\pi\epsilon_0 R_{ij}}. \tag{S13}$$

For parameters, $Q_{\mathrm{C}} = 0.6512$ $|e|$, and $Q_{\mathrm{O}} = -0.3256$ $|e|$ (where $e$ denotes the charge of the electron).

Given the CO$_2$ force field defined above, one can easily calculate the cavity-free force $\mathbf{F}_{nj}^{(0)}$ (see Eq. (S6)) as a function of the nuclear configurations by standard molecular dynamics packages like LAMMPS [11]. In CavMD, at each time step, we also need to evaluate the molecular dipole moment projected along direction $\boldsymbol{\xi}_\lambda = \mathbf{e}_x, \mathbf{e}_y$:

$$d_{ng,\lambda} = \left[Q_\mathrm{O}\left(\mathbf{R}_{n\mathrm{O}_1} + \mathbf{R}_{n\mathrm{O}_2}\right) + Q_\mathrm{C}\mathbf{R}_{n\mathrm{C}}\right] \cdot \boldsymbol{\xi}_\lambda \tag{S14}$$

and its derivative $\partial d_{ng,\lambda}/\partial \mathbf{R}_{nj}$, which is just the partial charge of each nucleus ($Q_\mathrm{O}$ or $Q_\mathrm{C}$). For the calculation of the IR spectrum (see below), the total dipole moment $\boldsymbol{\mu}_S$ projected along direction $\boldsymbol{\xi}_\lambda$ is calculated by $\boldsymbol{\mu}_S \cdot \boldsymbol{\xi}_\lambda = \sum_{n=1}^{N_\mathrm{sub}} d_{ng,\lambda}$.

*g. Spectroscopy of polaritons and dark modes* In the main text, the global infrared (IR) absorption spectrum is calculated by Fourier transforming the dipole auto-correlation function from equilibrium NVE trajectories [12–15]:

$$n(\omega)\alpha(\omega) = \frac{\pi\beta\omega^2}{2\epsilon_0 V c}\frac{1}{2\pi}\int_{-\infty}^{+\infty} dt\, e^{-i\omega t}\left\langle \sum_{i=x,y}(\boldsymbol{\mu}_S(0)\cdot\mathbf{e}_i)(\boldsymbol{\mu}_S(t)\cdot\mathbf{e}_i)\right\rangle. \tag{S15}$$

Here, $\alpha(\omega)$ denotes the absorption coefficient, $n(\omega)$ denotes the refractive index, $V$ is the volume of the system (i.e., the simulation cell), $\mathbf{e}_i$ denotes the unit vector along direction $i = x, y$, and $\boldsymbol{\mu}_S(t)$ denotes the *total* dipole moment of the molecules at time $t$, where $\boldsymbol{\mu}_S(t) = \sum_{nj} Q_{nj}\mathbf{R}_{nj}(t)$. See the above definitions around Eq. (S13) for the values of the partial charges $Q_{nj}$ of each nucleus (the true nuclear charge for nucleus $nj$ plus the electronic shielding effect), which is assumed to be a constant within the classical force field. In Eq. (S15), the summation over $x$ and $y$ is a summation over the two possible polarizations of the relevant cavity mode (where the cavity is assumed to be oriented along the $z$-direction; see Fig. 1 in the main text for the cavity setup).

When evaluating the nonequilibirum local IR spectroscopy for each molecule after the external pulse pumping, given a nonequilibrium NVE trajectory, we calculate

$$n(\omega)\alpha_{\mathrm{local},l}(\omega) = \frac{\pi\beta\omega^2}{2\epsilon_0 V c}\frac{1}{2\pi}\int_{-\infty}^{+\infty} dt\, e^{-i\omega t}\left\langle \boldsymbol{\mu}_l(0)\cdot\boldsymbol{\mu}_l(t)\right\rangle. \tag{S16}$$

Here, $\boldsymbol{\mu}_l(t) = \sum_j Q_{lj}\mathbf{R}_{lj}(t)$ denotes the dipole moment for molecule $l$.

Note that Eq. (S15) describes the collective behavior of the molecular dipole system, which reflects the polaritonic response under VSC. By contrast, Eq. (S16) provides information about the dynamics of each individual molecule, which is composed mostly of vibrational dark modes (since the molecular number in the simulation, $N_\mathrm{sub}$ is always 216 or more and the polaritonic contribution of each molecule is small).

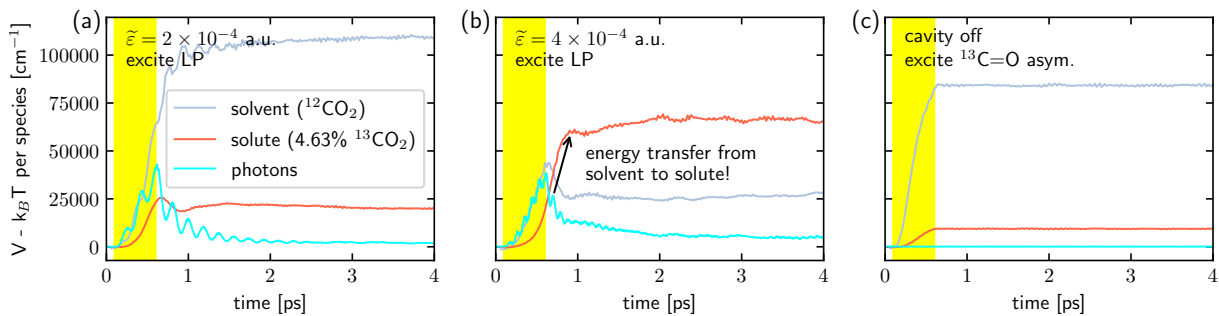

Supplementary Figure 2. Replot of the data in Fig. 3 of the manuscript except here the dynamics of the C=O bond energy per species (not per molecule) is shown. In other words, we plot the total amount of energy in the $^{12}CO_2$ and $^{13}CO_2$ species. The photonic potential energy is also plotted as the cyan line. As shown in Fig. (b), first the solvent (blue line) gains energy from interacting with the external field (i.e., during the yellow window); subsequently, after a small delay, the solvent transfers a large amount of energy to the solute (red line) after the field is turned off (i.e. after the yellow window). As the solvent polariton transfers energy to the solute molecules or solvent dark modes, the photonic energy is also reduced. This data shows unambiguously show that there is an ultrafast (sub-ps) vibrational energy transfer from the solvent to the solute molecules. Outside the cavity (Fig. (c)), such a fast intermolecular vibrational energy transfer is impossible, highlighting the benefits of forming vibrational polaritons and the use of a nonlinear mechanism to promote intermolecular vibrational energy transfer. To reduce signal fluctuations, we have applied a time-series moving average to the curves with neighboring four data points.

### SUPPLEMENTARY NOTE 1 | ULTRAFAST INTERMOLECULAR ENERGY TRANSFER FROM SOLVENT TO SOLUTE

In Supplementary Figure 2, we replot Fig. 3 in the manuscript but here we show the C=O bond energy per molecular species (not per molecule). In Supplementary Figure 2b, it is very clear that there is an ultrafast vibrational energy transfer from the solvent to the solute species, which is absent either outside the cavity (as in Supplementary Figure 2c) or when the solvent LP does not have a suitable frequency to support the nonlinear absorption of the solute molecules (as in Supplementary Figure 2a).

The timescale of intermolecular vibrational energy transfer in Supplementary Figure 2b is short which might lead the reviewer to question our interpretation of the dynamics above. In order to demonstrate the time-resolved energy transfer dynamics even more clearly, in

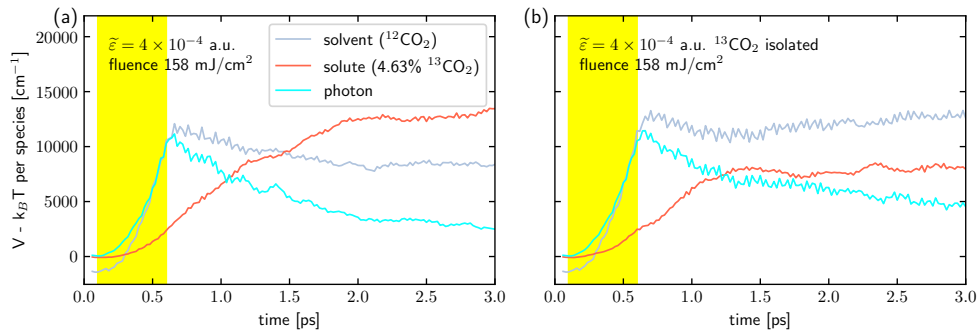

Supplementary Figure 3. (a) Similar plot as Supplementary Figure 2b except that the pulse fluence is reduced from 632 mJ/cm$^2$ to 158 mJ/cm$^2$. Under a smaller pulse fluence, after the pulse excitation (the yellow window), the energy transfer from the $^{12}CO_2$ solvent to the $^{13}CO_2$ solute species can be identified more easily. (b) Replot of Fig. (a) except that the intermolecular interactions involving $^{13}CO_2$ molecules are turned off. In the absence of the $^{13}CO_2$ intermolecular interactions, the $^{12}CO_2$ → $^{13}CO_2$ intermolecular vibrational energy transfer is greatly suppressed.

Supplementary Figure 3 we perform two additional simulations and plot the potential energy *per species*.

For the first simulation (to be compared against the simulation condition in Fig. 3b of the manuscript (or Supplementary Figure 2b)), in Supplementary Figure 3a we have reduced the pulse fluence from $F = 632$ mJ/cm$^2$ to $F = 158$ mJ/cm$^2$. Such a reduction of the pulse fluence can increase the polariton lifetime and give better time-resolved energy transfer dynamics than that in Supplementary Figure 2b. In Supplementary Figure 3a, after the pulse excitation (yellow window), we find that both the solvent and photonic energies are transferred to the solute molecules, showing that forming solvent polaritons can indeed facilitate intermolecular vibrational energy transfer. Although the solute concentration is only 4.63%, most the input energy is accumulated in the solute molecules at $t = 3$ ps.

For the second simulation, we further remove the intermolecular interactions involving the $^{13}CO_2$ species. In other words, Supplementary Figure 3b is an *unphysical* simulation where the intermolecular interactions between each $^{13}CO_2$ and other molecules ($^{12}CO_2$ or $^{13}CO_2$) are artificially turned off. Although Supplementary Figure 3b is unphysical, this plot shows unambiguously that, as compared with Supplementary Figure 3a, the energy transfer from the $^{12}CO_2$ species to the $^{13}CO_2$ species is greatly suppressed, demonstrating the importance of intermolecular interactions in facilitating energy transfer. Overall, Supplementary Figure 3b shows that our mechanism of energy accumulation in the solute species after the solvent

polariton pumping is indeed a many-body mechanism and intermolecular interactions play an important role.

## SUPPLEMENTARY NOTE 2 | EFFECT OF CAVITY LOSS

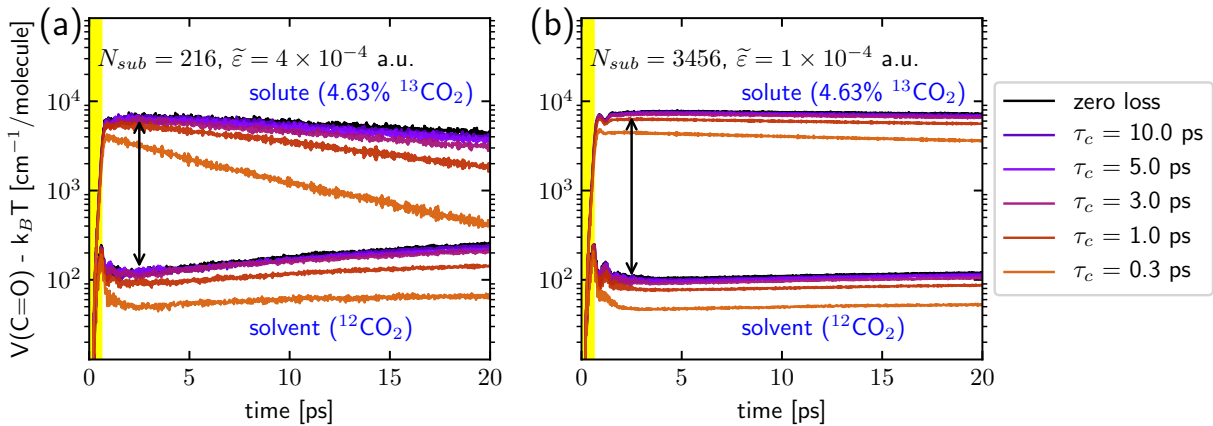

Supplementary Figure 4. (a) C=O bond energy dynamics per $^{13}CO_2$ solute and $^{12}CO_2$ solvent molecules when the molecular system size is $N_{\text{sub}} = 216$ and $\widetilde{\varepsilon} = 4 \times 10^{-4}$ a.u. (the same simulation parameters as Fig. 3b in the main text). Black to orange lines denote different cavity losses, ranging from zero loss to a cavity lifetime of 10.0, 5.0, 3.0, 1.0, and 0.3 ps. (b) The corresponding dynamics when the system size is enlarged by 16 times and the Rabi splitting is fixed (by taking $\widetilde{\varepsilon} = 1 \times 10^{-4}$ a.u.). These simulations indicate that, for Fabry–Pérot cavities (which corresponds to the limit when $N_{\text{sub}}$ is large), cavity loss does not meaningfully alter the reported dynamics in Fig. 3b once the cavity lifetime is larger than 1 ps.

The reported results in the main text did not include cavity loss. Here, in Supplementary Figure 4a, we rerun the main finding (Fig. 3b) in the main text by adding a cavity loss on top of the original simulation. All other parameters and simulation details are the same as in Fig. 3b. Color lines from black to orange denote the choice of different cavity loss (from zero cavity loss to a cavity lifetime of 10.0, 5.0, 3.0, 1.0, and 0.3 ps). The top lines represent the C=O bond energy dynamics per $^{13}CO_2$ solute molecule, while the bottom lines represent that of $^{12}CO_2$ solvent molecules. As shown in Supplementary Figure 4a, the simulation results are robust against the cavity loss when the cavity lifetime is longer (or the same as) 3 ps.

As discussed in Supplementary Methods, when using CavMD to simulate Fabry–Perot

microcavities, we need to enlarge the system size to check the convergence of the simulation results. Supplementary Figure 4b plots the corresponding C=O bond energy dynamics when the molecular number in the simulation cell is enlarged from $N_{\text{sub}} = 216$ to 3456 (by 16 times). Accordingly, in order to maintain the fixed Rabi splitting, the effectively light-matter coupling strength is reduced from $\widetilde{\varepsilon} = 4 \times 10^{-4}$ a.u. to $\widetilde{\varepsilon} = 1 \times 10^{-4}$ a.u. In this relatively large molecular system, cavity loss plays a role only when the cavity lifetime is smaller than 1 ps.

We can understand the cavity loss dependence of the dynamics as follows. The C=O bond dynamics can be divided into two phases: the polariton relaxation ($t < 1$ ps; since the polariton lifetime is 0.6 ps; see Fig. 2h in the main text) and later-time vibrational relaxation and energy transfer after the polariton decay ($t > 1$ ps). The former dynamics are dominated by the polariton, and the later dynamics are dominated by the dark modes. As far as the first phase (polariton relaxation) is considered, when the cavity loss is very large (or when the cavity lifetime is smaller than 1 ps), because cavity loss becomes an important channel for the energy dissipation, less polariton energy can be transferred (or dephased) to the vibrational dark modes, leading to a significantly altered C=O bond energy dynamics. By contrast, when the cavity lifetime becomes larger than 1 ps, the early-time dynamics is not altered by the cavity loss (for Supplementary Figure 4a,b). In later times (the second phase), for the dynamics of each individual molecule, since the polariton contribution is just $1/N_{\text{sub}}$ and the dark-mode dynamics dominate the molecular response, cavity loss no longer plays a meaningful role when $N_{\text{sub}}$ is large (Supplementary Figure 4b). For Supplementary Figure 4a, since $N_{\text{sub}}$ is not very large ($N_{\text{sub}} = 216$), cavity loss plays a more important role for the later-time dynamics than that in Supplementary Figure 4b ($N_{\text{sub}} = 3456$), with meaningful differences when the cavity lifetime is smaller than 3 ps.

Overall, when CavMD is used to simulate Fabry–Pérot microcavities, the system size enlarging process in Supplementary Figure 4 shows that cavity loss cannot meaningfully alter the reported dynamics when the cavity lifetime is greater than 1 ps (which is the usual parameter in Fabry–Perot cavities). Of course, Supplementary Figure 4 also indicates that, for small-volume cavities (such as plasmonic nanocavities) under VSC, a large cavity loss may suppress the selectivity of energy transfer.

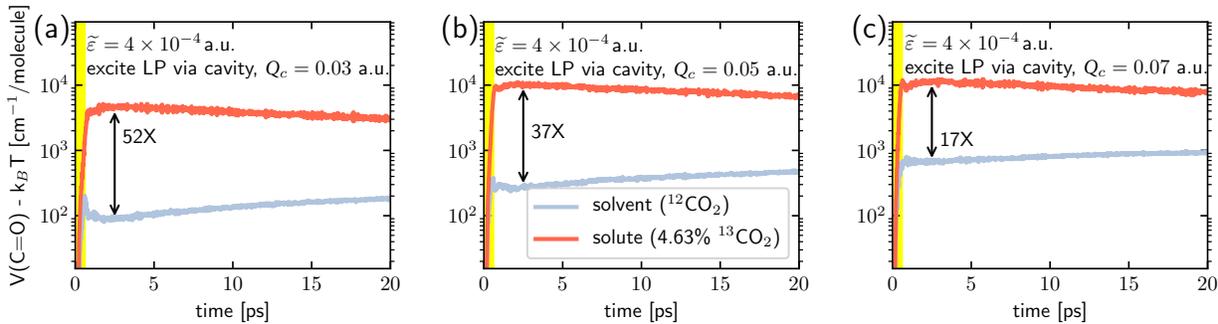

Supplementary Figure 5. Replot of Fig. 3b except for the change that the incoming field interacts only with the cavity mode via an effective charge $Q_c$ (rather than interacting with the molecular subsystem). All the other simulation details are the same as in Supplementary Figure 5b. In (a-c), we plot the C=O bond dynamics when $Q_c$ takes the values 0.03, 0.05, and 0.07 a.u., respectively. Note that similar amount of energy is infused to the coupled cavity-molecular system in both Fig. 3b (of the main text) and Fig. a (here).

### SUPPLEMENTARY NOTE 3 | EFFECT OF PUMPING CAVITY MODES

The reported results in the main text have assumed that the incoming field interacts with the molecular subsystem only. In Supplementary Figure 5, we replot Fig. 3b (a key figure in the main text) by assuming that the incoming field interacts with the cavity mode only via the following interaction form $-Q_c \tilde{\tilde{q}}_c \mathbf{E}_{\text{ext}}(t)$, where we have defined $Q_c$ as the effective charge of the cavity mode in Supplementary Methods. Keeping all the other simulation details are the same as in Fig. 3b, Supplementary Figure 5 plots the C=O bond dynamics for the solute and solvent molecules for different choices of $Q_c$: (from left to right) $Q_c =$ 0.03, 0.05, and 0.07 a.u. We have found that when $Q_c = 0.03$ a.u., Supplementary Figure 5a largely reproduces Fig. 3b, showing that exciting either the molecular or the cavity mode yields similar results when a similar amount of the input energy is infused to the coupled cavity-molecular system. Admittedly, for these simulations there is an ambiguity insofar as the fact that we must choose a value for $Q_c$; nevertheless, our conclusions are robust provided that $Q_c$ is chosen so that we dump the same amount of energy into the cavity in Supplementary Figure 5 as in Fig. 3b of the main text.

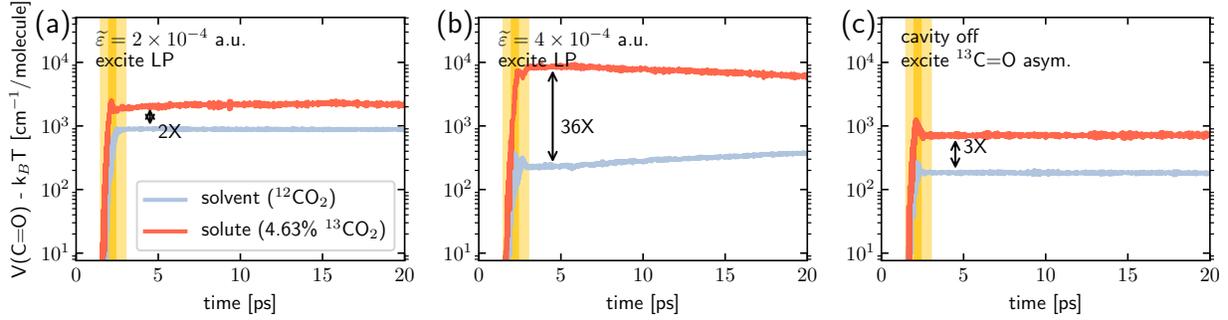

Supplementary Figure 6. Replot of Fig. 3 in the main text except for the use of an Gaussian pulse defined in Eq. (S17) instead of a continuous wave pulse (see Eq. (S8)). See text below for details.

### SUPPLEMENTARY NOTE 4 | EFFECT OF A GAUSSIAN PULSE

The reported results in the main text have assumed a pulse in the form of Eq. (S8). Here, we replace this pulse form by a Gaussian pulse:

$$\mathbf{E}_{\text{ext}}(t) = E_0 \exp\left[-2\ln(2)\frac{(t - t_0 - 4\tau_{\text{FWHM}})^2}{\tau_{\text{FWHM}}^2}\right]\cos(\omega t)\mathbf{e}_x, \quad (S17)$$

where we set $E_0 = 3.084 \times 10^7$ V/m ($6 \times 10^{-3}$ a.u.), $\tau_{\text{FWHM}} = 0.5$ ps, $t_0 = 0.01$ ps. In Supplementary Figure 6, we replot Fig. 3 in the main text for the case that we excite with this Gaussian pulse; all the other parameters are the same as in Fig. 3. The yellow vertical region denotes the time window when the molecular system experiences the peak of the Gaussian pulse. The results are largely the same as in Fig. 3. Note that, since the parameters for the Gaussian pulse have been chosen to resemble the wave pulse in Eq. (S8), a similar amount of energy is pumped into the coupled cavity-molecular system as in Fig. 3 of the main text.

### SUPPLEMENTARY NOTE 5 | EFFECT OF THE INPUT PULSE FLUENCE

Supplementary Figure 7 plots the dependence of selectivity (i.e., the C=O bond energy ratio between the $^{13}CO_2$ and $^{12}CO_2$ molecules at $t = 2.5$ ps) as a function of the incoming pulse fluence when the simulation parameters are identical to that in Fig. 3b except for the pulse fluence. The transient selectivity goes through a maximum as a function of the pulse fluence, confirming our assertion that energy transfer to the $^{13}CO_2$ species is dominated by a nonlinear transition: at low intensities this mechanism is inefficient, while at high intensities the $^{13}CO_2$ species energy absorption becomes saturated.

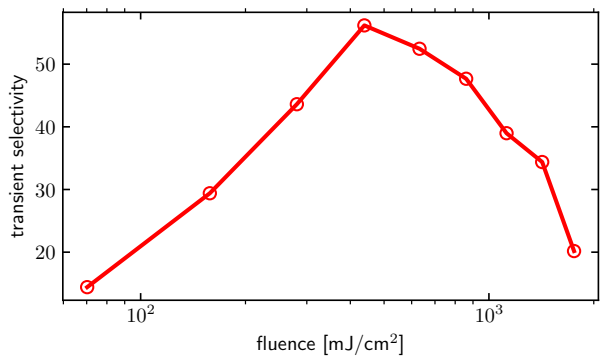

Supplementary Figure 7. C=O bond energy ratio between the average $^{13}CO_2$ and the average $^{12}CO_2$ molecule at $t = 2.5$ ps (i.e., the transient selectivity labeled in Fig. 3b) as a function of the input pulse fluence. Except for the change of pulse fluence, all the other parameters are the same as in Fig. 3b of the main text. Note that the transient selectivity goes through a maximum as a function of the pulse fluence. This inversion behavior comes from two competing effects: on the one hand, increasing the pulse fluence can enhance the nonlinear absorption for the $^{13}CO_2$ molecules which tends to increase the selectivity; one the other hand, enhancing the pulse fluence over a threshold saturates the energy absorption of all the molecules, which tends to reduce the selectivity.

### SUPPLEMENTARY NOTE 6 | EFFECT OF WEAK OR STRONG COUPLING

In Supplementary Figure 2 and Supplementary Figure 3, we have shown the uniqueness of our finding — under a suitable solvent polariton frequency, it is possible to achieve an ultrafast (sub-ps) and strong intermolecular vibrational energy transfer from the $^{12}CO_2$ solvent to the $^{13}CO_2$ solute molecules after pumping the solvent polariton. Here, we are interested in another aspect of our finding: under which conditions can we highly excite the few $^{13}CO_2$ solute molecules, especially do we need strong light-matter interactions?

It is clear that if the external field interacts only with the molecular bright mode rather than with the cavity mode, since forming a solvent polariton increases the effective dipole moment of the system, the presence of strong coupling will lead to more total absorption than one would find for the case of a few isolated $^{13}CO_2$ molecules in a cavity (with weak coupling). Then, if the energies of the solvent and solute allow for the nonlinear $0 \rightarrow 2$ nonlinear mechanism as described in the main body of the text, we might expect more energy accumulation in the $^{13}CO_2$ molecules (when dynamics proceed through a polariton

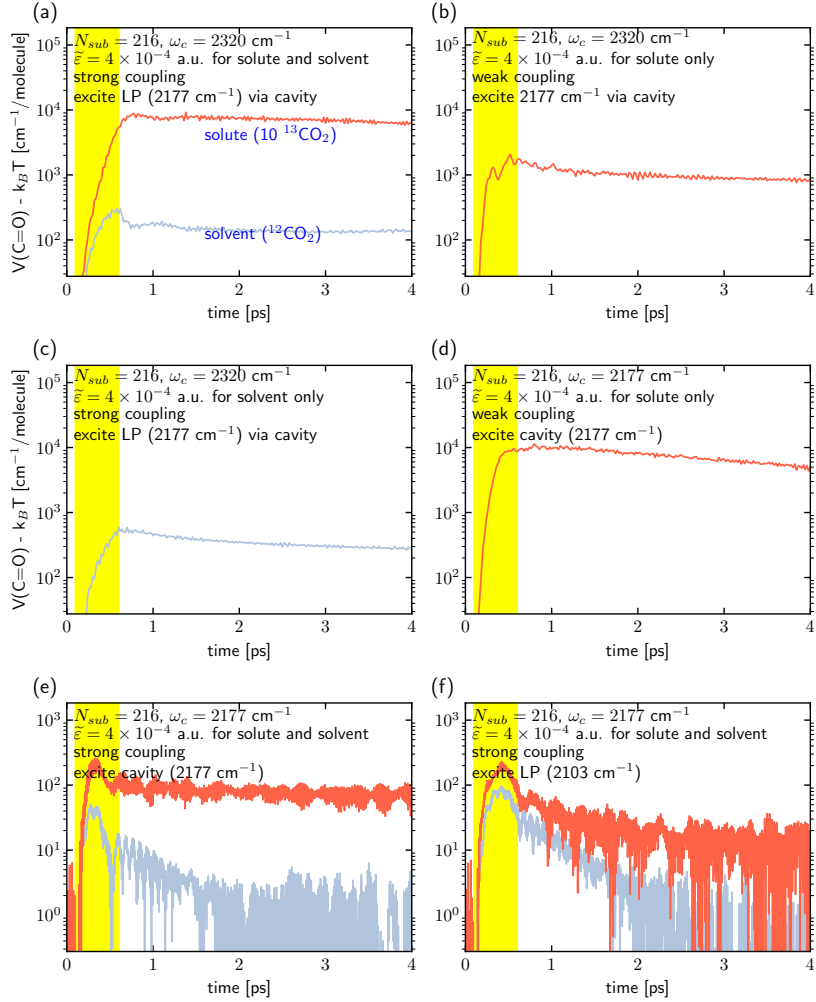

Supplementary Figure 8. C=O bond energy dynamics per $^{13}CO_2$ solute (red) or $^{12}CO_2$ solvent (gray) molecule when an external pulse peaked at 2177 cm$^{-1}$ excites the cavity mode (not molecules) with $Q_c = 0.05$ a.u. (a) Similar plot as Fig. 3b in the manuscript when the cavity mode ($\omega_c = 2320$ cm$^{-1}$) forms strong coupling with the solute/solvent system. (b) Cavity ($\omega_c = 2320$ cm$^{-1}$) contains the $^{13}CO_2$ species only. Other details are the same as in Fig. (a) (including the pulse excitation at 2177 cm$^{-1}$). Here, the solute is not strongly pumped because, without solvent, the incoming excitation is off-resonant. (c) Cavity ($\omega_c = 2320$ cm$^{-1}$) contains the $^{12}CO_2$ host only and the other details are the same as Fig. (a) (including the pulse excitation at the LP frequency 2177 cm$^{-1}$). The solvent energy absorption is larger than Fig. (a) because there is no solute molecules for energy accumulation. (d) Cavity ($\omega_c = 2177$ cm$^{-1}$) contains the $^{13}CO_2$ species only. The pulse excites at 2177 cm$^{-1}$ (the cavity mode frequency here). From Fig. (d), we find that, if we could turn off solvent-cavity interactions, we could simply build a cavity whose frequency is close the $^{13}CO_2$ frequency, pump at that same frequency, and excite the solute to a highly excited state without strong coupling. (e) Cavity ($\omega_c = 2177$ cm$^{-1}$) coupled to both the solute and solvent molecules and the pulse excites at 2177 cm$^{-1}$ (the cavity mode frequency here). (f) Similar plot as in Fig. (e) except that the external pulse frequency is set to 2103 cm$^{-1}$ (the LP frequency for this case). For Fig. (e), the solute is not strongly excited since the pumping is off-resonant because the presence of light-matter coupling $\widetilde{\varepsilon}$ shifts the energy of the cavity mode. For Fig (f), even when pumping the LP, we still find no enhancement of the $^{13}CO_2$ because there is no nonlinear mechanism (and energy match) to promote two quanta of excitation on a solute molecule. In Figs. (a)-(f), the cavity loss lifetime is set to $\tau_c = 0.4$ ps ($= 83$ cm$^{-1}$); this choice of $\tau_c$ is large enough such that, if the solvent were to interact with the cavity, there would be strong coupling. However, if only the solute interacts with the cavity, there is weak coupling — the Rabi splitting is less than the width of the cavity mode (as relevant to Figs. (b) and (d)) Altogether, in the presence of many $^{12}CO_2$ solvent molecules, there is no means of exciting the $^{13}CO_2$ molecules strongly and preferentially without illuminating the $^{12}CO_2$ solvent polariton and activating the nonlinear ($0 \rightarrow 2$) mechanism described in the main body of the text.

17

formed by strong coupling with the solvent molecule) than we would with simple excitation of a few $^{13}CO_2$ molecules (where there is no strong coupling at all). Indeed, this is what we have found.

One might wonder if the conclusions above are limited by the assumption that light interacts with the molecules directly (and not through the cavity). To that extent, we will now consider the parameter limit when only the cavity mode (not the molecules) interacts with the external pulse. We are interested in the situations when the few $^{13}CO_2$ molecules can be highly excited (no matter weak or strong coupling). In order to make a fair comparison, we have taken the light-matter coupling between the cavity mode and each individual molecule to be the same ($\widetilde{\varepsilon} = 4 \times 10^{-4}$ a.u.) throughout this comparison, meaning that the cavity volume is taken the same (i.e., we are interested in the same type of cavity setup).

We are interested in six different cases, which have been described in detail in the caption of Supplementary Figure 8. Supplementary Figure 8d shows that the $^{13}CO_2$ absorption can be enhanced in a suitably chosen cavity ($\omega_c = 2177$ cm$^{-1}$) under weak coupling if only a few $^{13}CO_2$ molecules are inside the cavity (i.e., we have manually removed the $^{12}CO_2$ solvent molecules). However, if we restrict ourselves to a molecular system with a large amount of $^{12}CO_2$ solvent molecules plus a few $^{13}CO_2$ solute molecules, Supplementary Figure 8e,f show even when the cavity frequency is tuned to $\omega_c = 2177$ cm$^{-1}$, because the presence of the solvent molecules will non-resonantly form a LP with the cavity mode (with $\omega_{\text{LP}} = 2103$ cm$^{-1}$), neither exciting the cavity mode nor exciting the LP frequency can lead to a strong excitation of the $^{13}CO_2$ molecules. Ultimately, as also shown in Supplementary Figure 3, to achieve a strong *selective* excitation of the $^{13}CO_2$ solute molecules in the presence of a bath fo $^{12}CO_2$ solvent molecules, the best approach is indeed to match the solvent mode with the cavity mode and excite the LP frequency — in other words, the approach prescribed in the main body of the present paper.

**SUPPLEMENTARY NOTE 7 | COMBINED EFFECTS OF CAVITY LOSS, PUMPING CAVITY MODES, AND A GAUSSIAN PULSE**

Lastly, let us consider the situation when an external Gaussian pulse pumps a lossy cavity mode directly. Such a parameter setting is similar to the input-output theory of optical cavities [16]. In this framework, the quantum coupling between an external electric

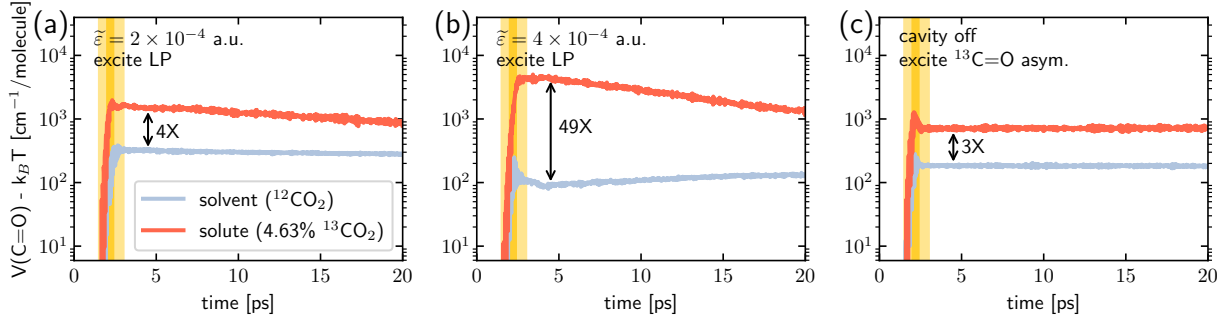

Supplementary Figure 9. Replot of Fig. 3 in the main text with the following differences: (i) an Gaussian pulse defined in Eq. (S17) is used instead of the wave pulse in Eq. (S8); (ii) in Figs. (a) and (b), the external field excites the cavity mode only with $Q_c = 0.028$ a.u. and the cavity lifetime is set as $\tau_c = 0.75$ ps. Again, we find large selective excitation of the solute molecules.

field and the cavity modes in a planar optical cavity can be defined as

$$\hat{H}_{\text{pump}} = i\hbar \int \frac{d^2\mathbf{k}_\|}{(2\pi)^2} \sum_\sigma \left[ \eta_\sigma^{\text{fr}}(\mathbf{k}_\|) \tilde{E}_\sigma^{\text{inc}}(\mathbf{k}_\|, t) \hat{a}_{c,\sigma}^\dagger(\mathbf{k}_\|) - \eta_\sigma^{\text{fr}*} \tilde{E}_\sigma^{\text{inc}*}(\mathbf{k}_\|, t) \hat{a}_{c,\sigma}(\mathbf{k}_\|) \right]. \quad (S18)$$

Here, $\hat{a}_{c,\sigma}^\dagger(\mathbf{k}_\|)$ and $\hat{a}_{c,\sigma}(\mathbf{k}_\|)$ denote the creation and annihilation operators of cavity photon with an in-plane wave vector $\mathbf{k}_\|$ and polarization direction $\sigma$, $\tilde{E}_\sigma^{\text{inc}}(\mathbf{k}_\|, t)$ denotes the Fourier transform of the incident field $\tilde{E}_\sigma^{\text{inc}}(\mathbf{r}, t)$, and $\eta_\sigma^{\text{fr}}(\mathbf{k}_\|)$ is the coefficient proportional to the transmission amplitude of the front mirrors of the planar optical cavity when the incident light has an in-plane wave vector $\mathbf{k}_\|$. The finite transmittivity of the front and back mirrors of the cavity leads to the transmitted and reflected electric fields, which take the following form:

$$\hat{E}_\sigma^{\text{tr}}(\mathbf{k}_\|, t) = \kappa_\sigma^{\text{back}}(\mathbf{k}_\|) \hat{a}_{c,\sigma}(\mathbf{k}_\|, t), \quad (S19a)$$

$$\hat{E}_\sigma^{\text{refl}}(\mathbf{k}_\|, t) = E_\sigma^{\text{inc}}(\mathbf{k}_\|, t) + \kappa_\sigma^{\text{fr}}(\mathbf{k}_\|) \hat{a}_{c,\sigma}(\mathbf{k}_\|, t), \quad (S19b)$$

where $\kappa_\sigma^{\text{fr,back}}(\mathbf{k}_\|)$ denotes the coefficient proportional to the transmission amplitude of the front or back mirrors. The cavity loss can be treated as a Lindblad form on top of the density matrix of the cavity field $\hat{\rho}$:

$$\mathcal{L}[\hat{\rho}] = \int \frac{d^2\mathbf{k}_\|}{(2\pi)^2} \frac{\gamma_\sigma^{\text{rad}}(\mathbf{k}_\|)}{2} \left[ 2\hat{a}_{c,\sigma}(\mathbf{k}_\|) \hat{\rho} \hat{a}_{c,\sigma}^\dagger(\mathbf{k}_\|) - \hat{a}_{c,\sigma}(\mathbf{k}_\|) \hat{a}_{c,\sigma}^\dagger(\mathbf{k}_\|) \hat{\rho} - \hat{\rho} \hat{a}_{c,\sigma}(\mathbf{k}_\|) \hat{a}_{c,\sigma}^\dagger(\mathbf{k}_\|) \right], \quad (S20)$$

where $\gamma_\sigma^{\text{rad}}(\mathbf{k}_\|)$ denotes the radiative decay rate of the cavity. In the follows the nonradiative decay rate of the cavity will be disregarded.

Under the simplest case when $\eta_\sigma^{\rm fr}(\mathbf{k}_\parallel) = \eta_\sigma^{\rm back}(\mathbf{k}_\parallel)$, $\kappa_\sigma^{\rm fr}(\mathbf{k}_\parallel) = \kappa_\sigma^{\rm back}(\mathbf{k}_\parallel)$, $\gamma_\sigma^{\rm rad}(\mathbf{k}_\parallel) = -2\eta_\sigma(\mathbf{k}_\parallel)\kappa_\sigma(\mathbf{k}_\parallel)$, and also assuming normal incidence $\mathbf{k}_\parallel = 0$, given the transmittivity of the cavity mirrors as $T \ll 1$, we have

$$\eta = \frac{cT}{2l_z}\sqrt{\frac{l_z}{\pi\hbar\omega_c}}, \tag{S21a}$$

$$\kappa = -T\sqrt{\frac{\pi\hbar\omega_c}{l_z}}, \tag{S21b}$$

$$\gamma = \frac{cT^2}{l_z}. \tag{S21c}$$

Let us connect the parameters in input-output theory to those (i.e., $\tau_c$ and $Q_c$) in CavMD. For a cavity mode of frequency $\omega_c = 2320$ cm$^{-1}$, the corresponding cavity length is $l_z = 2.16$ $\mu$m (with fundamental quantum number). Assuming $T = 0.025$, we compute the cavity lifetime as

$$\tau_c = \frac{1}{\gamma} = \frac{l_z}{cT^2} = 0.75 \text{ ps}. \tag{S22}$$

We further connect $\eta$ to $Q_c$. Using $\hat{a}_c^\dagger - \hat{a}_c = \sqrt{2\omega_c/\hbar}\hat{q}_c$ and the classical coupling as $-Q_c\tilde{\tilde{q}}_c\mathbf{E}_{\rm ext}(t)$ in CavMD, we may write

$$Q_c = \hbar\eta\sqrt{\frac{2\omega_c}{\hbar}} = \frac{cT}{\sqrt{2\pi l_z}} = 0.028 \text{ a.u.} \tag{S23}$$

Given the above values ($\tau_c = 0.75$ ps and $Q_c = 0.028$ a.u.), and also taking the Gaussian pulse defined above, when the other parameters are kept the same as Fig. 3 in the main text, we rerun CavMD simulations and plot Supplementary Figure 9 (in a similar way as Fig. 3 in the main text). Again, a large selectivity is observed inside the cavity.

### SUPPLEMENTARY NOTE 8 | A QUANTUM ILLUSTRATION OF OUR FINDING

The numerical findings reported in the main text can be justified by a quantum picture. As shown in the left box of Supplementary Figure 10, for an anharmonic potential surface with three energy levels $|0\rangle$ (the black rectangle), $|1\rangle$ (the red rectangle), and $|2\rangle$ (the yellow rectangle), when a strong pulse at the frequency which is half of the $|0\rangle \to |2\rangle$ transition pumps the molecular system, multiphoton nonlinear absorption can occur, which is facilitated by a virtual state (i.e., a second-order process in a Fermi's golden rule calculation).

Beyond this simple textbook picture of multiphoton nonlinear absorption, polaritons can also enhance the nonlinear absorption of molecules (see the middle box of Supplementary

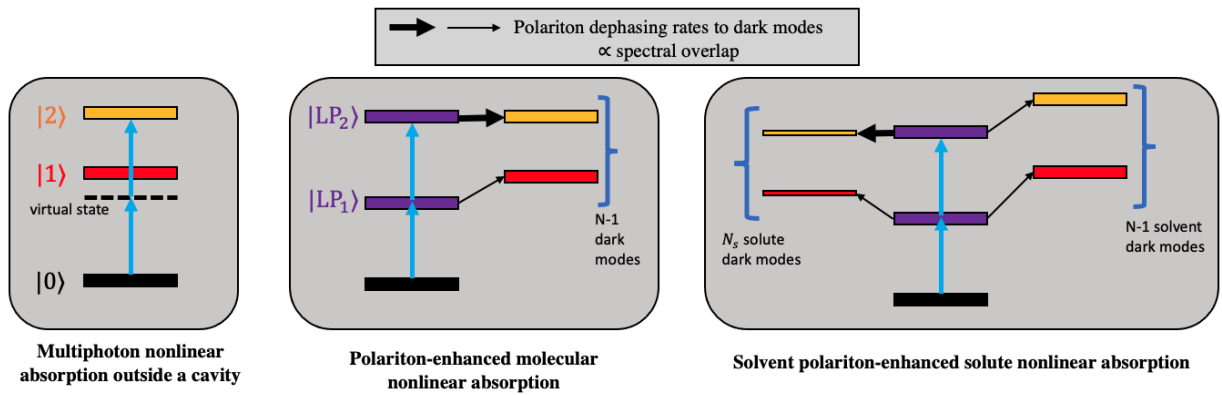

Supplementary Figure 10. A quantum illustration of selective exciting the solute molecules by pumping the solvent polariton. (left box) Illustration of multiphoton nonlinear absorption outside a cavity; (middle box) illustration of polariton-enhanced molecular nonlinear absorption; (right box) illustration of our finding in the main text, i.e., when the two LP state $|LP_2\rangle$ is near resonance with the second excited states of the solute dark modes, strongly pumping the LP may lead to a dominate dephasing channel from the LP to the second excited state of a set of solute molecules. The black arrows denote the polaritonic dephasing channels to the dark modes and the thickness of the arrows represents the rate of the dephasing rate (due to spectral overlap).

Figure 10). Considering the case when $N$ identical anharmonic molecules are coupled to a cavity mode and strong coupling is formed, the first excited state of the LP ($|LP_1\rangle$, the purple rectangle) can dephase to the $N-1$ first excited states of the dark modes ($|1_d\rangle$). Similarly, the second excited state of the LP ($|LP_2\rangle$, the purple rectangle) can dephase to the second excited states of the dark modes ($|2_d\rangle$). A Fermi's golden rule calculation shows that the polaritonic dephasing rate is proportional to the spectral overlap between the initial (polaritonic) and the final (dark) states. Now, when $|LP_2\rangle$ is near resonance with $|2_d\rangle$, the large spectral overlap between the $|LP_2\rangle$ and $|2_d\rangle$ states will lead to a very fast dephasing process for the $|LP_2\rangle$ state. When the incoming field is strong enough, $|LP_2\rangle$ can be meaningfully populated and outperform the dephasing of the $|LP_1\rangle$ state, which leads to polariton-enhanced molecular nonlinear absorption.

Finally, as shown in the right box of Supplementary Figure 10, we consider the case when $N$ solvent molecules form strong coupling with a cavity mode and $|LP_2\rangle$ has a very small spectral overlap with the solvent $|2_d\rangle$ states. In this case, polariton-enhanced molecular nonlinear absorption is greatly suppressed for the solvent molecules. Now, if we add a small

concentration of the solute molecules to the system, the presence of the solute molecules will not alter the frequencies of the polaritons. If the $|2_d\rangle$ states of the solute molecules (the thin yellow rectangle) are near resonance with the $|\text{LP}_2\rangle$, a strong excitation of the LP will result in enhanced nonlinear absorption for the solute molecules only.

In principle, one can construct master equations to describe the above mechanism. However, from our perspective, such a master equation would depend critically on the four dephasing rates ($|\text{LP}_1\rangle$ or $|\text{LP}_2\rangle$ to the solute or solvent degrees of freedom) and the reliability of any such master equation would be questionable without *ab initio* modeling (ideally quantum modeling) — which is currently not practical. In this sense, quantum master equations may not be more realistic than our fully atomistic classical CavMD simulations. For this reason, we have chosen not to propagate the master equations. Nevertheless, the quantum illustration in Supplementary Figure 10 should help understand our findings from a quantum perspective.

---

[S1] T. E. Li, J. E. Subotnik, and A. Nitzan, Cavity molecular dynamics simulations of liquid water under vibrational ultrastrong coupling, Proc. Natl. Acad. Sci. **117**, 18324 (2020).

[S2] T. E. Li, A. Nitzan, and J. E. Subotnik, Cavity molecular dynamics simulations of vibrational polariton-enhanced molecular nonlinear absorption, J. Chem. Phys. **154**, 094124 (2021).

[S3] C. Cohen-Tannoudji, J. Dupont-Roc, and G. Grynberg, *Photons and Atoms: Introduction to Quantum Electrodynamics* (Wiley, New York, 1997) pp. 280–295.

[S4] T. E. Li, A. Nitzan, and J. E. Subotnik, On the origin of ground-state vacuum-field catalysis: Equilibrium consideration, J. Chem. Phys. **152**, 234107 (2020).

[S5] T. S. Haugland, E. Ronca, E. F. Kjønstad, A. Rubio, and H. Koch, Coupled Cluster Theory for Molecular Polaritons: Changing Ground and Excited States, Phys. Rev. X **10**, 041043 (2020).

[S6] J. Flick, M. Ruggenthaler, H. Appel, and A. Rubio, Atoms and Molecules in Cavities, from Weak to Strong Coupling in Quantum-Electrodynamics (QED) Chemistry, Proc. Natl. Acad. Sci. **114**, 3026 (2017).

[S7] T. E. Li, A. Nitzan, and J. E. Subotnik, Collective vibrational strong coupling effects on molecular vibrational relaxation and energy transfer: Numerical insights via cavity molecular dynamics simulations, Angew. Chemie Int. Ed. , anie.202103920 (2021).

[S8] L. Martínez, R. Andrade, E. G. Birgin, and J. M. Martínez, PACKMOL: A package for building initial configurations for molecular dynamics simulations, J. Comput. Chem. **30**, 2157 (2009).

[S9] R. T. Cygan, V. N. Romanov, and E. M. Myshakin, Molecular Simulation of Carbon Dioxide Capture by Montmorillonite Using an Accurate and Flexible Force Field, J. Phys. Chem. C **116**, 13079 (2012).

[S10] B. d. Darwent, *Bond dissociation energies in simple molecules* (U.S. National Bureau of Standards, 1970).

[S11] S. Plimpton, Fast Parallel Algorithms for Short-Range Molecular Dynamics, J. Comput. Phys. **117**, 1 (1995).

[S12] D. A. McQuarrie, *Statistical Mechanics* (Harper-Collins Publish- ers, New York, 1976).

[S13] M.-P. Gaigeot and M. Sprik, Ab Initio Molecular Dynamics Computation of the Infrared Spectrum of Aqueous Uracil, J. Phys. Chem. B **107**, 10344 (2003).

[S14] S. Habershon, G. S. Fanourgakis, and D. E. Manolopoulos, Comparison of path integral molecular dynamics methods for the infrared absorption spectrum of liquid water, J. Chem. Phys. **129**, 074501 (2008).

[S15] A. Nitzan, *Chemical Dynamics in Condensed Phases: Relaxation, Transfer and Reactions in Condensed Molecular Systems* (Oxford University Press, New York, 2006).

[S16] I. Carusotto and C. Ciuti, Quantum fluids of light, Rev. Mod. Phys. **85**, 299 (2013).



\end{document}